\documentclass[useAMS,usenatbib]{mn2e}
\usepackage{graphics,epsfig,psfig}
\usepackage[normalem]{ulem}
\usepackage{xcolor}
\usepackage[]{inputenc,amssymb}

\def \be{\begin{equation}}
\def \ee{\end{equation}}
\def \bea{\begin{eqnarray}}
\def \eea{\end{eqnarray}}
\def\etal{{et al.\ }}

\def\ltsima{$\; \buildrel < \over \sim \;$}
\def\simlt{\lower.5ex\hbox{\ltsima}}
\def\gtsima{$\; \buildrel > \over \sim \;$}
\def\simgt{\lower.5ex\hbox{\gtsima}}
\newcommand{\angstrom}{\textup{\AA}}

\definecolor{webgreen}{rgb}{0,.5,0}
\definecolor{webbrown}{rgb}{.6,0,0}
\usepackage[pdfpagelabels]{hyperref}
\hypersetup{%
   colorlinks=true,hyperfootnotes=false,%
   breaklinks=true,%
   plainpages=false, bookmarksnumbered, bookmarksopen=true,
   bookmarksopenlevel=1,%
   urlcolor=webbrown, linkcolor=webbrown, citecolor=webgreen,
   }

\title[Molecular Outflows]{
Molecular outflows in starburst nuclei}

\author[Arpita Roy, Biman B. Nath, Prateek Sharma, Yuri Shchekinov]
{Arpita Roy$^{1,2}$ \thanks{arpita@rri.res.in}, Biman B. Nath$^1$, Prateek Sharma$^2$, Yuri Shchekinov$^3$\\
$^1$Raman Research Institute, Sadashiva Nagar, Bangalore 560080, India\\
$^2$Joint Astronomy Programme and Department of Physics, Indian Institute of Science, Bangalore 560012, India\\
$^3$ P. N. Lebedev Physical Institute, 53 Leninskiy Prospekt, 119991 Moscow, Russia\\
}

\begin{document}

\maketitle

\label{firstpage}

\begin{abstract}
Recent observations have detected molecular outflows in a few nearby starburst nuclei. We discuss the 
physical processes at work in such an environment in order to outline a scenario that can explain the 
observed parameters of the phenomenon, such as the molecular mass, speed and 
size of the outflows. 
We show that outflows triggered by OB associations, with $N_{OB}\ge 10^5$ (corresponding to a star formation 
rate (SFR)$\ge 1$  M$_{\odot}$ yr$^{-1}$ in the 
nuclear region), in a stratified disk with mid-plane 
density $n_0\sim 200\hbox{--}1000$ cm$^{-3}$ and scale height $z_0\ge 200 (n_0/10^2 
\, {\rm cm}^{-3})^{-3/5}$ pc, can  form molecules in a cool dense and expanding shell. The associated molecular 
mass is $\ge 10^7$ M$_\odot$ at a 
distance of a few hundred pc, with a speed
of several tens of km s$^{-1}$. We show that a SFR surface density of $10 \le \Sigma_{SFR} \le 50$ M$_\odot$ yr$^{-1}$
kpc$^{-2}$ favours the production of molecular outflows, consistent with observed values.
%We also speculate that subsequent evolution of the outflowing gas can give rise to warm molecules, such as 
%observed in M82 at about $\sim 1$ kpc, 
%when the outflowing shell 
%stalls at a few times the scale height. 
\end{abstract}


\section{Introduction}
\label{intro}
Observations show that outflows from starburst galaxies contain gas in different phases, which manifest 
with different emission mechanisms and are probed in different
wavelengths. The fully ionised component usually show up through 
free-free emission and is probed by X-ray observations (\cite{strickland2004}, \cite{heckman1990}). Partially 
ionized/atomic component are more clumpy than the fully ionised gas, 
and are probed by line emission from various ions, e.g. NaI, MgII etc \citep{heckman00}. 
Outflows from some nearby starburst galaxies have also been observed 
to contain a molecular component. Understanding the dynamics of this 
molecular component has become an important issue, in light of recent 
observations with {\it ALMA} and further observations in the future. 

\cite{bolatto2013} observed a molecular outflow in the central region of NGC 253 with a rate of $\ge 3 $ 
M$_\odot$ yr$^{-1}$ (likely as large as 
$9$ M$_\odot$ yr$^{-1}$), with a mass loading factor $1\hbox{--}3$. 
Four expanding shells with radii $60\hbox{--}90$ pc have velocities of $\simeq 23\hbox{--}42$ 
km s$^{-1}$, suggesting a dynamical age of $\sim 1.4\hbox{--}4$ Myr. The inferred 
molecular mass is $(0.3-1) \times 10^7$ M$_\odot$, %with a momentum $\sim (8.5-40)\times 10^7$ 
%M$_\odot$ km s$^{-1}$, 
and energy $\sim (2-20)\times 10^{52}$ erg. These shells likely outline
a larger shell around the central starburst region.

\cite{tsai2012} observed a molecular outflow in NGC 3628 with the CO (J=1-0) line. 
The outflow shows almost a structureless morphology with a very weak bubble breaking through 
in the north part of the central outflow.
Its size of $\sim 370\hbox{--}450$ pc, inferred molecular mass of $\sim 2.8 \times 10^7$ M$_\odot$, and outflow speed 
$\sim 90\pm 10$ km s$^{-1}$, suggest a total kinetic energy of molecular gas of $\sim 3 \times 10^{54}$ erg. 

More recently \citep{salak2016} observed dust lanes above the galactic plane in NGC 1808 along with NaI, NII, CO(1-0) emission lines tracing extraplanar gas close (within 2 kpc) to the galactic centre with a mass of $10^8~M_\odot$, and a nuclear star formation rate of $\sim 1~M_\odot$ yr$^{-1}$. The velocity along the minor axes varies in the range $48\hbox{--}128$ km s$^{-1}$ and most likely indicates a gas outflow off the disk with an estimated mass loss rate of $(1\hbox{--}10)~M_\odot$ yr$^{-1}$.

The molecular outflow observed in M82 has a complex morphology. The part of it outlined by CO emission
is at a larger radii than the part seen with HCN and HCO$^+$ lines. The CO (J=1-0) observations show
diffuse molecular gas in a nearly spherical region of radius $\sim 0.75$ kpc, with
a total molecular mass $3.3 \times 10^8$ M$_\odot$, with an average outflow velocity 
of $\sim 100$ km s$^{-1}$ \citep{walter2002}. The corresponding kinetic energy of the CO-outflow is 
of $\sim 3\times 10^{55}$ erg. %The outflow 
%shows diffuse distribution without signs of a shell-like structure expected from a presumably 
%large scale central explosion that had initiated the outflow. Currently it covers a nearly spherical 
%region of 1.5 kpc in diameter. 
More recently \cite{salak2013} re-estimated the mass and kinetic energy of CO gas to be larger
 by factors of 3 and 3-10 respectively.
Notably, the molecular outflow morphology is similar to that of the 
dust halo described by \cite{alton1999}.

The morphology of the region of the outflow observed in HCN/HCO$^+$ is similar 
to that of the CO outflow -- it is amorphous and nearly spherical with 
slightly smaller length scale: the radius of the HCN region is of $400 \hbox{--}450$ pc, and around $600$ pc for HCO$^+$; 
both HCN and HCO$^+$ emissions show clumpy structure with characteristic size of 100 pc \citep{salas2014}. 
The kinematics and the energetics differ slightly from those inferred for the
CO-outflow: the mean de-projected outflow velocity for HCO$^+$ is 64 km s$^{-1}$, while for HCN it 
is 43 km s$^{-1}$. The total molecular mass contained into the HCN (HCO$^+$)
outflows is 
$\ge 7 (21)\times 10^6$ M$_\odot$, which in total 
is an order of magnitude lower for molecular outflows associated with CO \citep{walter2002}. The kinetic energy 
of the outflow associated with HCN/HCO$^+$ emission ranges between $5\hbox{--}30 \times 10^{52}$ 
erg. The molecular outflow rate is $\ge 0.3$ M$_\odot$ yr$^{-1}$. 
They also inferred 
a SFR of $\sim 4\hbox{--}7$ M$_\odot$ yr$^{-1}$ from free-free emission. %, which is 5 to 8 times 

These observations pose a number of questions that we address in this paper: are the molecules formed {\it in situ} in the flow or are
they entrained the flow, or are the residues of the parent molecular cloud (in which the superbubble has gone off)?
What are the typical length scales, time scales, molecular mass and speed? How are these related to the SFR, or 
disk parameters (e.g., gas density, scale height)? 

In this paper we outline a model which includes the basic physical processes for producing a molecular
outflow in starburst nuclei, and addresses some of these issues. We have kept our model simple enough
to be general, but it has the essential ingredients in order to explain some of the observed parameters
mentioned above, namely the length scales and velocities, as well as an estimate of the molecular mass. 
Our results can become the base models of more sophisticated numerical simulations
which would be able to address finer details of this complex phenomenon.

We use a model of a shell propagating in a stratified ISM in our calculation. Such an outflow is inherently 2-dimensional,
with the dense shell pushed out to a roughly constant stand-off radius in the plane of the disk, while the top of the bubble is
blown of by Rayleigh-Taylor instability. In steady state, a dense shell (in which molecules can form) exists in a dynamically 
young ($r/v\sim $ few Myr) conical shell confined within a few times the scale-height (see Figure \ref{fig:schem} for a cartoon;
for numerical simulations, see Figs 2, 3 of \cite{sarkar2015}). For analytical tractability, we consider the formation 
and survival of molecules in the dense shell expanding in a stratified disk. All starburst nuclei discussed in the paper show
a CO disk and biconical outflows emanating from them. We expect our simple estimates to apply, at least to an order of magnitude,
for the realistic scenario.

We begin with a discussion of the phase space of molecular and ionic components of outflows from
starbursts, and after eliminating various possibilities we arrive at a basic scenario (\S 2). In the later
part of this section, we study various physical constraints on the parameters of the starburst and the
disk galaxy for producing molecular outflows. Next we discuss the physical processes involved in 
the formation and destruction of molecules in these outflows (\S 3) and present our results in \S 4.

\section{Arriving at a physical model}
\subsection{Radius-velocity space of molecular and atomic components}
The phase space of the outflows with molecular and ionic components that we 
introduce below can be instructive in order to arrive at a physical model.
Consider the case of a dense shell of a superbubble triggered by an OB association. In the case of a 
uniform ambient medium density $\rho$, the position and velocity of the shell are given by,
\be
r \sim \Bigl ( {\mathcal{L} t^3 \over \rho} \Bigr ) ^{1/5} \,;\quad v \sim {3 \over 5} \Bigl ( {\mathcal{L} \over \rho} \Bigr ) ^{1/5} t^{-2/5} \approx {3 \over 5} \Bigl ( {\mathcal{L} \over \rho} \Bigr ) ^{1/3} r^{-2/3} \,,
\ee
where $\mathcal{L}$ is the mechanical luminosity driving the superbubble. 
In other words, the position and 
velocity of the shell are related as $r \propto v^{-3/2}$. We can first compare this with observational data. 
However, in order to make a meaningful comparison between galaxies with different SFR, one can take out 
the dependence on SFR, by plotting $v/({\rm SFR})^{1/5}$ 
against $r/({\rm SFR})^{1/5}$, since both $r$ and $v$ depend on $\mathcal{L}$ (and consequently, the SFR) 

\begin{figure}
\centerline{
\epsfxsize=0.52\textwidth
\epsfbox{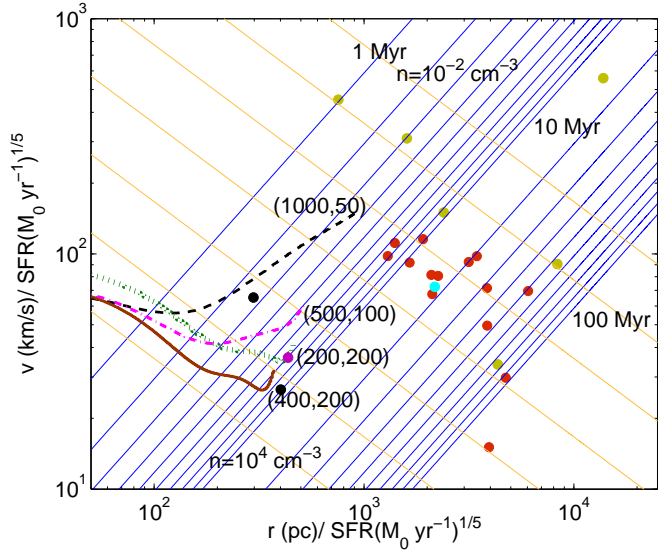}
} 
\caption{
Phase space of molecular and atomic outflows, with points representing different observations of molecular 
(black  and magenta points) and 
atomic outflows (olive green points), as well as atomic outflows from ULIRGs (red points). 
The cyan point represent the warm (2000 K) molecular outflow of M82. %and the olive green point represent the atomic outflow in M82. 
The black-dashed, 
green-dotted, magenta-dashed-dotted and the brown solid lines show the simulation 
results for superbubble evolution with radiative cooling for different combinations 
of mid-plane density  and scale height (as labelled, with the first number of the
pair being density in cm$^{-3}$ and the second being the scale height in pc). Orange solid lines represent 
the $v\textendash r$ lines 
for different fixed hydrogen 
particle densities (of the ambient medium) ranging from $0.01$ cm$^{-3}$ (top) to $10^4$ cm$^{-3}$ (bottom), and for a given mechanical luminosity injection. The density increases from top to bottom with the 
increment by a factor 
of 10 between two consecutive lines. The blue solid lines are for  different epochs in the logarithmic scale. 
The first ten lines are 
separated by 1 Myr starting from 1 Myr to 10 Myr, and the rest of the ten lines have a separation of 10 Myr between two consecutive lines ranging from 10 Myr to 
100 Myr.
}
\label{fig:mol_atom_comp_obs_sim_plot}
\end{figure}

We show in Figure \ref{fig:mol_atom_comp_obs_sim_plot} data from observations of  molecular components from NGC 253
and NGC 3828 (black 
points), and M82 (HCN component in magenta, and the warm component in cyan). The length scales and velocities of this component are known from imaging and spectroscopic observations. 
It is not easy to determine the distances of atomic clouds that are usually probed by absorption 
lines. However there are a handful of 
cases of outflows from edge-on galaxies where one has reliable information on the position and 
speed of the atomic clouds 
(M82, NGC 3079, 5253, 1482, 4666, 1808). These are shown as olive green points in the same figure. 
We also show the data of atomic components 
in outflows from ULIRG as red points (from \cite{martin2005}).

%%%%%%%%%%%%%%%%%%%%%%%%%%%%%%%%%%%%%%%%%%%%%%%%%%%%%%%%%%%%%%%%%%%%%%%%%%%%%%%%%%%%%%%%%%%%%%%%%%%%%%%%%
%%%%%%%%%%%%%%%%%%%%%%%%%%%%%%%%%%%%%%%%%%%%%%%%%%%%%%%%%%%%%%%%%%%%%%%%%%%%%%%%%%%%%%%%%%%%%

 We also show as brown slanted lines the simple scaling of $v \propto r^{-2/3}$, for different uniform 
ambient densities 
 $\rho$, with the density being the largest in the bottom-left corner ($n_0=10^4$ cm$^{-3}$) and 
the smallest  ($n_0=10^{-2}$ cm$^{-3}$) in the top-right corner of the diagram. The blue lines 
are the iso-chrones at different times, starting with 1 Myr from the top and each line 
separated by 1 Myr from the next, and separated by 10 Myr after the 10 Myr mark.

It is interesting that the molecular and atomic components separate out into different regions in the phase-space. 
(One cyan point among the olive green and red points for atomic 
component refers to the case of warm (2000 K) molecules in M82.) They also separate 
out with regard to the constant density lines although there are some exceptions.  In other words, 
molecules appear to probe small scale outflows and high ambient density 
regions, whereas the atomic components probe large scale outflows ($\ge 1$ kpc) and low ambient density. 
However, we note that here the length scales and velocities
are normalized by ${\rm SFR}^{1/5}$ and so the diagram may not allow such a neat interpretation
in terms of length scales and velocities.

The molecular and atomic components may not appear to be parts of an evolutionary sequence in the 
context of a uniform ambient medium, 
but they may be related if the density is not uniform as in the case of a stratified disk gas. 
We show the evolution of the vertical heights of 
superbubbles triggered by an association of $10^5$ OB stars (the reasons for this choice of 
parameter will be explained later), in an exponentially stratified medium
characterised by mid-plane density $n_0$ and scale height $z_0$. The dashed, dot-dashed, 
dotted and solid lines show the cases for different combinations of mid-plane
density and scale heights (labelled by these parameters, $n_0$ in units of cm$^{-3}$, and $z_0$ in pc). 
The evolutionary tracks are different from slanted lines because
of stratification and radiative cooling in the simulations. 
However, the point to be noted is that the 
atomic/ionic outflow data points may be connected to the molecular outflow data points through such 
evolutionary curves of superbubbles in a stratified medium, connecting these two apparently disparate phenomena with an 
evolutionary sequence.

\subsection{Preliminary estimates}
These observations lead to a few preliminary estimates. For example,
%\tcg{Very preliminary conclusions can be drawn from these facts: 
from typical sizes and velocities in observed outflows in NGC 253 and NGC 3068 one 
infers a dynamical age 
of $r/v \sim 2\hbox{--}4$ Myr. Assuming that the age of the star cluster associated with the
outflow is longer than the main sequence lifetime of the least massive OB star, {\it i.e.} 30 Myr, and with a 
constant SFR of a few M$_\odot$ yr$^{-1}$, 
the total %\sout{number of 
 number of SNe exploded %}\tcb{kinetic energy deposited} 
during the dynamical time scale is $\sim %\tcb{equivalent to the number to SNe exploded as} 
10^5$, and a total SNe energy of $\sim 10^{56}$ erg. The total kinetic energy deposited by these
SNe is $\sim 7 \times 10^{55}$ erg. However, when a SN remnant enters the radiative phase, its
energy (both kinetic and thermal) is lost and a small fraction remains in the form of kinetic energy.
\citep{smith1993} have shown that the energy of SN remnants decrease as $\propto R^{-2}$ in the radiative phase. 
Assuming that SN remnants merge with each other earlier than when their radii grow 3 times since the onset of radiative 
phase \citep{nath2013}, we arrive at the estimate of kinetic energy available for molecular outflow as $\simeq 7\times 10^{54}$ 
erg. Therefore the observed
 kinetic energy of the molecular outflow ($<10^{53}$ erg s$^{-1}$) 
is much smaller than the mechanical energy deposited by stars. 
This is even smaller than the mechanical energy retained by the superbubble, 
assuming that 90\% of the mechanical energy is lost via radiative cooling
\citep{sharma2013, vasiliev2015, gupta2016}.

These considerations point towards the following scenario.
Suppose that the central starburst drives a shell by multiple SNe 
explosions. A quasi-spherical expanding shock wave from an OB-association 
becomes unstable against Rayleigh-Taylor and Kelvin-Helmholtz instabilities 
(the latter begins to be operative when the shock front expands into the halo where the 
front goes up faster due to decreasing density  and a tangential component 
emerges) \citep{mmml}. In the next stage the uppermost part of the front breaks 
and forms an outflow in the vertical direction, while the rest of the shell 
fragments and forms multiple clouds and clumps moving pervasively within the 
expanding shell. The expanding molecular gas can be swept up at the observed 
distance $D\sim 500$ pc by the quasi-spherical shock wave propagating in an 
exponentially stratified gas layer with the scale height $z_0=100$ pc and the 
mid-plane density $n_0=3\times 10^2$ cm$^{-3}$, such that characteristic 
cooling time at $T\sim 10^6$ K is only 100 yr, and the current observed state 
of molecular outflow is consistent with the fact the majority of energy has 
been lost. 

We elaborate on this model in the rest of the paper. However, let us
consider here briefly the possibility that the molecular clumps are pushed by radiation pressure.
Molecular clumps are dense enough to ensure tight collisional coupling between dust and gas particles. 
In such conditions the radiation force acting on the clump, and the resulting acceleration, are
\be 
F_R\sim {\pi R^2\over c}\Phi, \qquad\qquad a_R\sim {3\over 8}{\Phi\over m_{\rm H}N_{\rm H}c},  
\ee
where $R$ is the clump radius, $\Phi$ is radiation energy flux, $N_{\rm H}$ is the 
column density of the clump, where we explicitly assumed 
$N_{\rm H_2}=2N_{\rm H}$. The energy flux can be estimated as 
$\Phi\sim 300 N_\ast L_\odot/4\pi D^2$, for a Kroupa IMF with $N_\ast$ being total number of stars 
in the underlying central stellar association, $D$ is the distance of clumps 
from the galactic  centre. For the same IMF one can assume 
$N_\ast\simeq 100N_{OB}$, $N_{OB}$ being the number of OB stars in the 
association. For $D\sim 500$ pc, $N_{\rm H}\simgt 10^{22}$ 
cm$^{-3}$ for a typical molecular cloud, one obtains 
\be \label{upv}
v\sim\sqrt{2\int a_R dr} %\simlt 500\sqrt{N_{OB}}~{\rm cm~s}^{-1}
\sim 30~{\rm km~s^{-1}} \times (N_{OB}/10^5)^{1/2}\,.
\ee
Therefore radiation pressure alone cannot possibly explain the typical length
scale and velocities of the observed outflows. Moreover, 
although the molecular outflow is dynamically
young, the nuclear starburst may be old enough such that most luminous O-stars 
(producing radiative
acceleration) are absent. 1-D numerical simulations show that radiative acceleration 
plays a subdominant role after a few Myr \citep{gupta2016}.

%%%%%%%%%%%%%%%%%%%%%%%%%%%%%%%%%%%%%%%%%%%%%%%%%%%%%%%%%%%%%%%%%%%%%%%%%%%%%%%%%%%%%%%%%%%%%%%%%%%%%%%%%%
%%%%%%%%%%%%%%%%%%%%%%%%%%%%%%%%%%
\begin{figure}
\centerline{
\epsfxsize=0.52\textwidth
\epsfbox{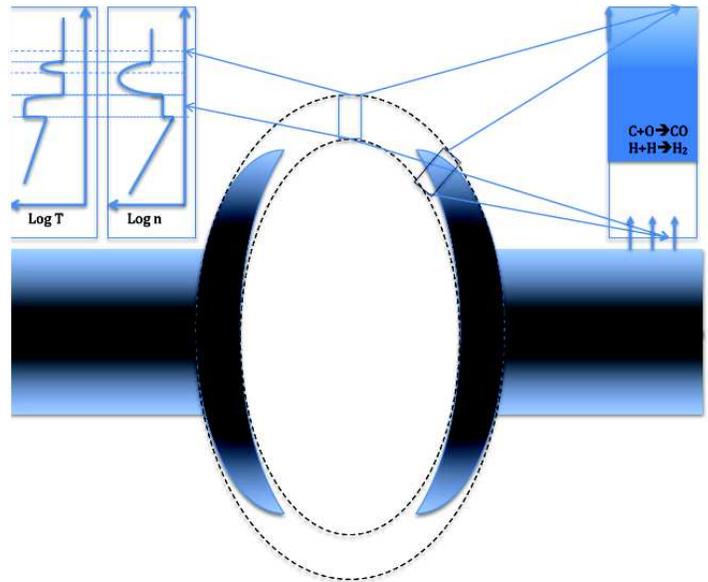}
}
\caption{Schematic diagram for the model of outflow used in this paper, with a superbubble 
shell ploughing through a stratified disk. The observed morphology is shown in grey tones, 
and the idealised superbubble shell is shown with dashed lines. A zoomed version of the shell
is shown on the right, highlighting the region where CO forms (for details, see \S 4.2). The arrows at the bottom of the
zoomed shell denote photons incident on the shell. Another zoomed version of the shell is 
shown on the left, that portrays the density and temperature profile in and around the shell.
See \S 4.1 for an explanation of this aspect.
}
\label{fig:schem}
\end{figure}
%%%%%%%%%%%%%%%%%%%%%%%%%%%%%%%%%%%%%%%%%%%%%%%%%%%%%%%%%%%%%%%%%%%%%%%%%%%%%%%%

%\subsection{Lower limit to molecular outflow length scale}
\section{Physical model}
\label{n0_z0_limit}
Consider a central OB association embedded in a dense stratified disk. The stratification in the disk 
is assumed to be exponential, with
a scale height of $z_0$ and a mid-plane  particle density
$n_0$. The ambient temperature  is assumed to be in the range of few tens of K, 
appropriate for a dense region, with densities in the 
range of $100\hbox{--}1000$ cm$^{-3}$. The mechanical luminosity arising from stellar processes 
in the OB association drives a shock
through the ambient medium, and this superbubble sweeps up ambient matter, which cools and 
forms a dense shell. The morphology of
the observed molecular outflows (mentioned in \S 1) suggests an epoch when the shell has broken 
out of the disk, as shown in the schematic
diagram in Figure \ref{fig:schem}. The observed morphology is shown in grey tones whereas the
idealised scenario of a superbubble adopted in this paper is shown with dashed lines. \cite{mmml} have
shown that most of the mass in the superbubble is confined to regions near the disk. 
However, for
analytical simplicity of a 1-dimensional calculation, we assume a quasi-spherical shell, and consider
its height as the indicator of its distance. The diagram also shows a zoomed version of the shell,
and highlights the region of CO formation which will be our region of interest for the calculation
of molecule formation/destruction (\S 4.2). Another schematic zoomed view of the shell is shown
on the left that shows the density and temperature profiles in and around the shell. We will describe
this structure in detail in \S 4.1.

The parameters used for the 
stratified disk (the midplane density from 100 to 10$^3$ cm$^{-3}$, and the 
scale height 50 to 200 pc) correspond roughly to 
a molecular cloud with the surface gas density of $\Sigma_g\simeq (10^3-10^4)M_\odot$ pc$^{-2}$ (equivalent to typical  
column densities of molecular clouds $N({\rm H})=10^{23}\hbox{--}10^{24}$ cm$^{-2}$), or the 
total mass of molecular cloud with size $D=1$ kpc of $M\sim 2\times 10^8-10^{10}M_\odot$ (for $\mu=2$, where $\mu$ is the mean molecular 
weight).

Consider the minimum size of the OB association needed to explain the observations
which show outflowing shell at $\sim 500$ pc with a speed $\sim 50$ km/s. The observations
of a molecular mass of $\ge 10^7$ M$_\odot$ implies a minimum gas density of $\sim 100$ cm$^{-3}$.
For a superbubble expanding in a uniform medium of density $\rho$, the required mechanical luminosity 
$\mathcal{L}$
for an outflow to have a speed $v$ at distance $r$, is given by $
\mathcal{L} \approx (5/3)^3 \rho \, v^3 r^2$.
The above mentioned observed parameters of speed, distance and density, therefore implies a minimum mechanical
luminosity of $\ge 10^{41}$ erg s$^{-1}$, which corresponds to $\ge 10^5$ OB stars (which we refer 
to as $N_{OB}$) \cite{roy2013}. %\sout{ exploding as SNe over $30$ Myr. }
We will use this value of $N_{OB}$ as a fiducial number in our work here.

%%%%%%%%%%%%%%%%%%%%%%%%%%%%%%%%%%%%%%%%%%%%%%%%%%%%%%%%%%%%%%%%%%%%%%%%%%%%%%%%%%
\begin{figure}
\centerline{
\epsfxsize=0.52\textwidth
\epsfbox{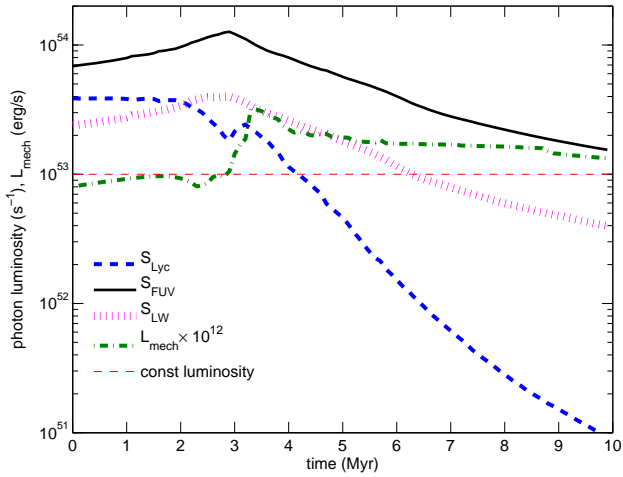}
}
\caption{The evolution of mechanical luminosity ($L_{mech}$), Lyman continuum photon luminosity and luminosity 
in the FUV ($S_{\rm FUV}$), and Lyman-Werner band for $N_{OB}=10^5$ ($S_{\rm LW}$), calculated using Starburst99. 
In this figure, we have plotted 
$L_{mech}\times 10^{12}$ to accommodate the mechanical luminosity curve along with the other luminosity plots.
The slowly growing part on mechanical luminosity on initial stages ($t<2$ Myr) is due to 
active stellar wind from massive stars; at $t>3$ Myr SNe explosions become dominant. }
\label{fig:lum_comp}
\end{figure}
%%%%%%%%%%%%%%%%%%%%%%%%%%%%%%%%%%%%%%%%%%%%%%%%%%%%%%%%%%%%%%%%%%%%%%%%%%%%%%%%

The evolution of the Lyman continuum luminosity of the central source is 
calculated with Starburst99 code for instantaneous star formation. We also show 
the evolution of Lyc and Lyman-Werner band photon luminosities in Figure 
\ref{fig:lum_comp} for $N_{\rm OB}=10^5$. Although the mechanical luminosity varies 
with time,
for simplicity we use a constant value ($10^{41}$ erg s$^{-1}$) as shown by a red 
dashed horizontal line, and which is a reasonable
approximation within the time scale of 10 Myr considered here. Note the initial rise and subsequent decline
in the FUV luminosity (solid line) with time. This behaviour of the FUV luminosity will be important in 
understanding the evolution of the thermodynamics of the shell, as will be described later in the paper.

\subsection{A flow-chart of our calculation strategy}
We first describe the formalism of our calculation before discussing the details. 
\begin{itemize}
\item {\it Dynamics} :--We first study the dynamics of a superbubble in a stratified medium. 
Since the density is large, cooling is important, and therefore
the standard solution of \cite{weaver1977} for uniform media, or the Kompaneets 
approximation \citep{kompaneets1960} for an adiabatic shock wave from a point explosion in a stratified atmosphere, is not adequate.
We use hydrodynamical simulations with gas cooling in order to obtain the evolution of the shell. 
However, in order to focus on the essential
physical processes, we only use the vertical height of the shell (denoted here by $z_+$), and 
ignore the effects of deviations from sphericity of the shell. We describe
the numerical set up in \S \ref{num}, and use the results of $z_+(t)$ and the corresponding 
superbubble velocity in our calculation.

\item {\it Thermodynamics}:-- Given the knowledge of the dynamics of the superbubble, we 
then discuss the (density and temperature) structure of the
shocked gas in \S \ref{2_zone}. We focus on thermodynamics of the the cool, dense shell that 
forms behind the shock. We assume the density jump between the shell density and the stratified ISM density
to be constant in time, for simplicity.
Although not precise, this assumption allows us to glean qualitative trends.
The estimates of the density jumps are given in \ref{2_zone} in the presence of the 
ISM magnetic field. Then we describe the dominant heating 
processes (photo-electric heating) and gas 
cooling (\S \ref{heat}). The photo-electric (PE) heating rate is 
calculated for the FUV photon 
luminosity ($S_{FUV}$) with dust extinction (see Appendix \ref{PE} for details). 
One also needs to have an estimate of the electron density (n$_e$) to calculate the PE-heating rate. 
The diffuse ISM UV photon luminosity is responsible for ionization of the ambient gas, that lies
outside the Str\"omgren sphere for the central source. We calculate 
$n_e$ assuming ionization equilibrium. We solve two coupled equations using 
thermal equilibrium (equating PE-heating with cooling) and ionization 
equilibrium to obtain the equilibrium shell temperature (T$_{shell} (t)$), and n$_e$.
 We also demonstrate that the
heating and cooling 
time scales are much shorter than the dynamical time scale, 
%\sout{and we argue that gas in this dense 
%shell quickly settles down to an equilibrium temperature} 
and thus thermal equilibrium in the shell is 
a good approximation (\S \ref{thermal_equil}).
\item {\it Molecule formation}:--Equipped with the knowledge of the density and temperature of 
the dense shell, we discuss the processes of molecule formation
and destruction (using H$_2$ as a proxy for all molecules) in \S \ref{mol_form_destruction}, and 
calculate the amount of molecules formed in the shell in different cases (\S \ref{mol_net}). The 
Lyman-Werner band photon luminosity (S$_{LW}$)  is used to calculate the 
photo-dissociation of the molecules, after taking into account the effect of dust extinction.
\end{itemize}

\begin{figure}
\centerline{
\epsfxsize=0.54\textwidth
\epsfbox{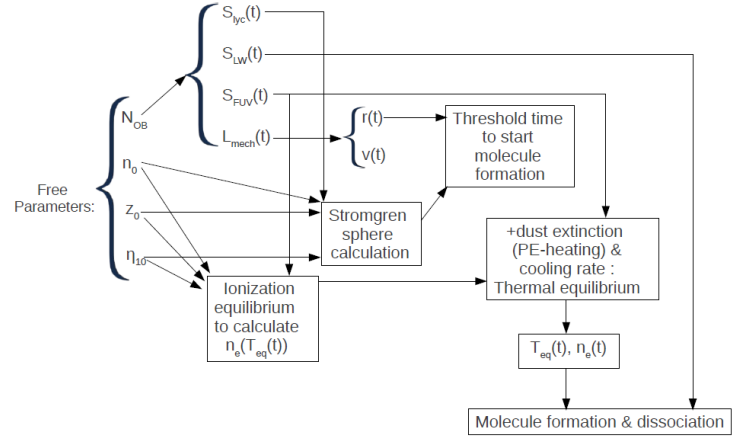}
}
\caption{The schematic diagram of the flowchart of the calculation}
\label{fig:flowchart}
\end{figure}

A schematic diagram of the flowchart of the calculation strategy is shown in figure \ref{fig:flowchart}.

%%%%%%%%%%%%%%%%%%%%%%%%%%%%%%%%%%%%%%%%%%%%%%%%%%%%%%%%%%%%%%%%%%%%%%%%%%%%%%%%%%%%%%%%%%%%%%%%
%%%%%%%%%%%%%%%%%%%%%%%%%%%%%%%%%%%%%%%%%%%%%%%%%%%%%%%%%%%%%%%%%%%%%%%%%%%%%%%%%%%%%%%%%%%%%%

However, the formation of molecules requires some basic conditions to be met. 
In section \ref{thresh}, we discuss the 
threshold conditions for molecule formation
in the outflowing shell, determined by the ionization due to 
the OB association.

\subsection{Threshold conditions for molecule formation in outflows}
\label{thresh}
When the OB association is born, the initial spurt of ionizing photons will 
send an ionisation front propagating through the surrounding medium, asymptotically forming an 
ionized zone (Str\"omgren sphere).  The gas will be largely swept-up in a shell such that the remaining gas becomes as dilute as 
having 2 to 2.5 orders of magnitude 
lower density than in the host cloud \citep{freyer2003,garcia2013,dale2014}. 
At the same time the supernovae and stellar winds arising in the OB association trigger an expanding superbubble that 
ploughs through the surrounding medium. 
Conditions inside the ionisation front will not support the  formation any molecules, and any existing 
molecule (entrained from the parent molecular cloud) will likely get 
photo-dissociated.
Therefore, as long as the superbubble shell is inside the 
ionisation front, its shell will propagate in a low-density environment and will not show any molecules. 
Initially, the ionization front would always move faster than the superbubble, and 
molecules cannot form in the shell 
until it has overtaken the ionization front. 
%The size of the Str\"omgren sphere therefore provides a lower limit to the 
%size of molecular outflows.

\begin{figure}
\centerline{
\epsfxsize=0.52\textwidth
\epsfbox{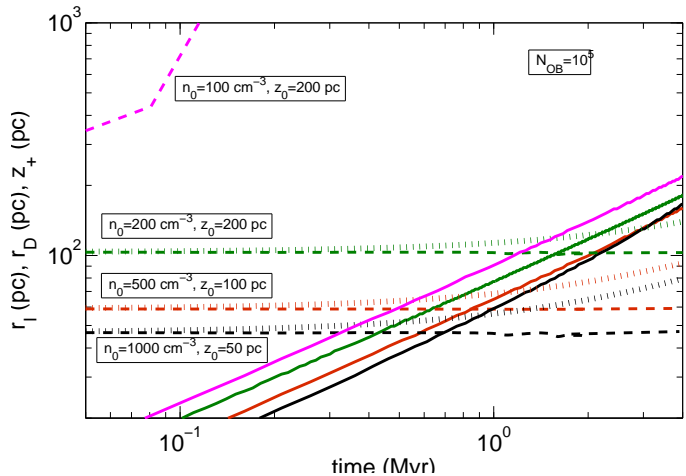}
}
\caption{The evolution of the ionisation front (the dotted and dashed lines) and the 
superbubble shell (solid lines), for $N_{OB}=10^5$ for four sets of n$_0$-z$_0$. 
The black lines represent the 
maximum density case (n$_0=1000$ cm$^{-3}$, $z_0=50$ pc), the red, and the green lines represent
n$_0=500$ cm$^{-3}$, z$_0=100$ pc, and n$_0=200$ cm$^{-3}$, z$_0=200$ pc respectively. 
The magenta curve refers to the case of $n_0=100$ cm$^{-3}$, $z_0=200$ pc. The dashed lines represent the Str\"omgren radii for the ambient medium with exponential density stratification, and time varying LyC photon luminosity for the corresponding sets of $n_0\textendash z_0$. The dotted lines represent the D-type ionisation front for the corresponding $n_0\textendash z_0$ cases.}
\label{fig:ion_front}
\end{figure}

We use hydrodynamical simulation (described in detail in Appendix A) in order to calculate the evolution 
of the superbubble. The mechanical luminosity driving the 
superbubble is assumed to be
constant in time,
$\mathcal{L} \approx 10^{41}$ erg s$^{-1} \times (N_{OB}/10^5)$ as obtained from Starburst99. 
In order to focus our attention 
to the basic physical processes, we consider the evolution of the vertical height ($z_+$) of bubble and compare with the ionization front in that direction.

The evolution of the ionization front $r_{I}$ is obtained by integrating,
\begin{equation}
 {dr_I \over dt}={{(S_{Lyc}(t)-4\pi r_{I}^3 \alpha_{HI} n^2/3)} \over {4\pi r_{I}^2 n}}\,,
\label{eq:rI}
\end{equation}
where $\alpha_{HI}$ is the case B recombination coefficient of hydrogen. We have calculated $r_I$ for the Lyc photon luminosity 
profile ($S_{Lyc} (t)$) as obtained from Starburst99, and for an
exponentially stratified density structure.
As the ionisation front propagates, it gives rise to a D-type front, whose distance can be estimated by eqn 37.26 of \cite{draine2011}. The epoch of conversion in to a D-type front can be estimated by eqn 37.15 of \cite{draine2011}. 

For simplicity, we have calculated the evolution of the ionization front and the superbubble shell independently.
In other words, the shell is assumed to move in a neutral medium, and the ionization front is assumed to move
in a uniform density medium.
For most cases, except for $n_0=100$ cm$^{-3}$, $z_0=200$ pc, the corresponding error is small, because the gas mass
within the Str\"omgren sphere is not large enough to considerably change the dynamics of the shell.

Figure \ref{fig:ion_front} shows the evolution of the ionization front, and the vertical location of the superbubble shell 
for four sets of n$_0\textendash$z$_0$ values and for N$_{\rm OB}=10^5$. 
The figure shows that
the St\"omgren sphere radius in the different cases is of order $\sim 50\hbox{--}100$ pc (shown by the black, red and green
dashed lines). The
ionization fronts transform to D-type fronts (prior to $0.1$ Myr) and expand slowly beyond the Str\"omgren radii, to reach
heights of order $\sim 80\hbox{--}140$ pc in $\sim 4$ Myr. At the same time, the corresponding superbubble shells overtake the ionization
fronts at heights of order $\sim 60\hbox{--}120$ pc.

Figure \ref{fig:ion_front} also shows that there are two different regimes: at low 
column densities of the layer $N({\rm H})=n_0z_0$  the ionization front moves ahead so quickly  such 
that the superbubble shell can never catch up with, and the shell propagates in a low-density ionized gas (as shown
by magenta lines). At
high column density limit, the shell can overtake the ionization front,
as shown by black, red and green solid lines. 

\begin{figure}
\centerline{
\epsfxsize=0.50\textwidth
\epsfbox{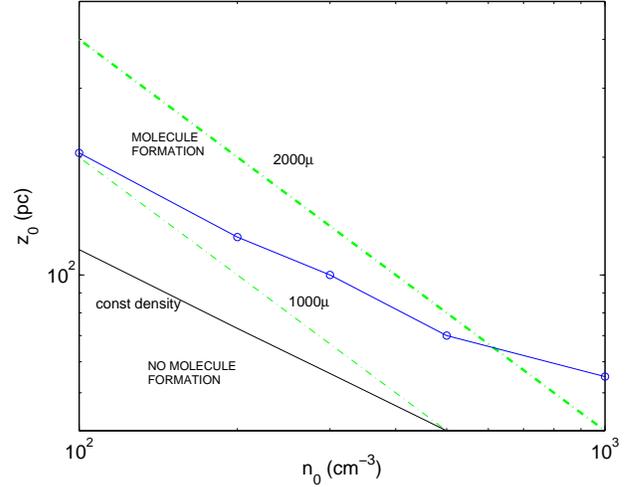}
}
\caption{The threshold combination of mid-plane density and scale height for the formation
of molecules in an outflowing shell triggered by an OB association with $N_{OB}=10^5$.  
The blue-solid line represent the cut-off $n_0\textendash z_0$ condition below which molecules can not form.
%\tcr{The threshold condition shown here corresponds to Ly-continuum and LW photons during 10 Myr. 
The green dashed-dotted lines correspond to two values of constant surface densities in units of $M_\odot/{\rm pc}^2$, where $\mu$ is the 
mean molecular weight. The black line plots the Str\"omgren radii for ambient medium with uniform densities for comparison.
}
\label{fig:threshold}
\end{figure}

It is reasonable to contend that molecular outflows can form behind the supernova-driven shock wave
only in the latter case.
These considerations lead us to determine the locus of the threshold
combination of $n_0$ and $z_0$ for the formation of molecules in outflowing shell. 
We show the result with the thin blue solid line in Figure 
\ref{fig:threshold}. The curve can be approximated by a fit, 
%roughly corresponding 
%to the Str\"omgren radius:
\be
z_0 \ge 200 \, {\rm pc} \, \Bigl ( { n_0 \over 10^2 \, {\rm cm}^{-3}} \Bigr )^{-3/5}
\ee
Note that for a uniform ambient medium of density $n_0$, and a constant 
Lyc luminosity of 
$4 \times 10^{53}$ photons s$^{-1}$,
 the Str\"omgren radius is given by $\sim 116 \, {\rm pc} (n_0/10^2 \, {\rm cm}^{-3})^{-2/3}$.
The difference between this estimate
and the above fit is due to (a) density stratification and (b) 
variation of Lyc luminosity with time. 

In Figure \ref{fig:threshold} we also show lines of constant surface density, marked in 
the units of M$_\odot$ pc$^{-2}$. We find that
molecular outflow is possible in a starburst nuclei region with surface density roughly 
$\ge 1500\mu \, ({\rm SFR}/{\rm M}_\odot \, {\rm yr}^{-1}) $ 
M$_\odot$ pc$^{-2}$.

It is interesting that \cite{nath2013} derived a lower limit on the molecular 
surface density of order $1000$ M$_\odot$ pc$^{-2}$
in starburst nuclei for producing outflows. 
A surface gas density of $\Sigma\sim 1.5 \times 10^3 \, \mu$ M$_\odot$ pc$^{-2}$ with $\mu=1.33$ implies
a SFR surface density of $\sim 10$ M$_\odot$ yr$^{-1}$ kpc$^{-2}$ from Kennicutt-Schmidt law \citep{kennicutt1998}.
This is incidentally $\sim 100$ times larger than the threshold SFR surface density for galactic superwinds \cite{heckman2015}.
Considering a nucleus region of radius $\sim 300$ pc, this translates to a SFR of $\ge 3$ M$_\odot$ yr$^{-1}$.

At the same time, our threshold relation also puts an upper limit on the SFR that can produce molecular outflows.
Recall that the size of the Str\"omgren sphere depends on the
ratio of $(L_{\rm LyC}/n^2)^{1/3}$. Since the density stratification has rendered a scaling of $z_0 \propto n^{-3/5}$,
therefore our threshold condition on the scale height likely scales as
$N_{\rm OB}^{3/10}$, or $({\rm SFR})^{3/10}$. Recalling that $N_{\rm OB}=10^5$ corresponds to
a SFR of $\sim 0.3$ M$_\odot$ yr$^{-1}$, we can re-write our threshold condition as, 
\be
z_0 \ge 200 \, {\rm pc} \, \Bigl ( { n_0 \over 10^2 \, {\rm cm}^{-3} }\Bigr )^{-3/5} \, \Bigl ( {{\rm SFR} \over 0.3 \, M_\odot \, {\rm yr}^{-1}} 
\Bigr )^{3/10} \,
\ee
Considering the size of the central starburst nuclei region to be $\sim 300$ pc, we can transform this
relation to one involving surface densities. We have,
\bea
\Sigma_{\rm SFR} && \le 50 \, M_\odot \, {\rm yr}^{-1} \, {\rm kpc}^{-2}  \nonumber\\
&&\times  \Bigl ( {\Sigma \over  1.5\times 10^3 \, \mu \, M_\odot \, {\rm pc}^{-2}} \Bigr )^{10/3} \, \Bigl ( { n_0 \over 10^2 \, {\rm cm}^{-3}} \Bigr )^{-4/3} 
\eea
This essentially implies that the SFR has to be lower than a certain value for a given column density of the starburst nucleus;
a larger SFR than the above inequality would inhibit the formation of molecules by ionizing the gas in the superbubble shell.

We then have three relevant scales for SFR surface density. A lower limit of $\Sigma_{SFR} \ge 0.1$ M$_\odot$ yr$^{-1}$ kpc$^{-2}$
ensures a galactic wind. However, the production of molecular outflows is limited to SFR surface densities $10 \le \Sigma_{SFR} \le 50  $ M$_\odot$ yr$^{-1}$ kpc$^{-2}$. 
It is interesting to note that the SFR surface densities of galaxies observed to host molecular outflows fall in this range. Therefore
our threshold condition for disk parameters for hosting molecular outflows is consistent with observations. 
We should, however, emphasize that galaxies show a considerable variation around the Kennicutt-Schmidt law and the above constraint on SFR surface density may not be a strong one.

%%%%%%%%%%%%%%%%%%%%%%%%%%%%%%%%%%%%%%%%%%%%%%%%%%%%%%%%%%%%%%%%%%%%%%%%%%%%%%%%%%%%%%%%%%%%%%%%%%%%
%%%%%%%%%%%%%%%%%%%%%%%%%%%%%%%%%%%%%%%%%%%%%%%%%%%%%%%%%%%%%%%%%%%%%%%%%%%%%%%%

\section{Shell density and temperature}
\label{shell_temp}
Given the knowledge of the threshold density and scale height for molecule formation, the next important
issues for molecules to form are the heating and cooling in the superbubble shell.
In this section, we discuss the heating and cooling processes that play an important role in determining the 
shell temperature. 

%%%%%%%%%%%%%%%%%%%%%%%%%%%%%%%%%%%%%%%%%%%%%%%%%%%%%%%%%%%%%%%%%%%%%%%%%%%%%%%%%%%%%%%%%%%%%%%%%%%%%%%%%
\begin{figure*}
\centerline{
\epsfxsize=1.0\textwidth
\epsfbox{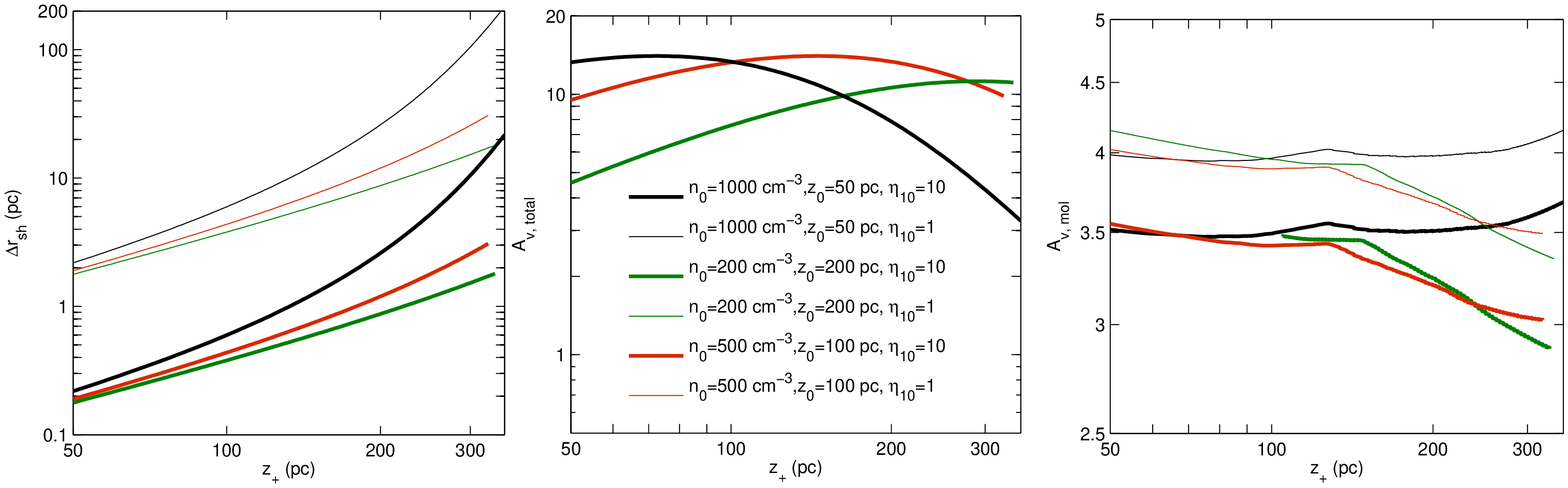}
}
\caption{The evolution of the shell thickness and A$_v$ for 
three mid-plane density and scale height combinations for $N_{OB}=10^5$, and for 
for three different $n_0$, $z_0$ cases ($n_0=1000$ cm$^{-3}$, $z_0=50$ pc; $n_0=200$ cm$^{-3}$, $z_0=200$ pc, and $n_0=500$ cm$^{-3}$, $z_0=100$ pc), for two $\eta_{10}$ 
cases
and for $N_{OB}=10^5$. 
The thick, and thin solid lines correspond to $\eta_{10}=10$, $1$ respectively. The black, green, and the red lines represent $n_0=1000$ cm$^{-3}$, $z_0=50$ pc; $n_0=200$ cm$^{-3}$, $z_0=200$ pc, and $n_0=500$ cm$^{-3}$, $z_0=100$ pc cases respectively. All the plots hereafter follow the same colour and line-styles for the corresponding 
$n_0$, $z_0$, and $\eta_{10}$ values. 
The left-most panel shows the shell thickness as a 
function of the vertical position of superbubble shell, and the middle panel represents the evolution of total 
A$_v$ (which does not depend on the value of $\eta_{10}$, or in other words, the thin lines coincide with the thick ones). The right panel
shows the values of $A_v ({\rm mol})$ for the region where molecules form in substantial quantity, and which is the region of our concern. 
 }
\label{fig:delr_Av_zp}
\end{figure*}
%%%%%%%%%%%%%%%%%%%%%%%%%%%%%%%%%%%%%%%%%%%%%%%%%%%%%%%%%%%%%%%%%%%%%%%%%%%%%%%%%%%%%%%%%%%%%%%%%%%%%%%

\subsection{Four-zone structure}
\label{2_zone}
From the point of view of the presence of molecular gas in the expanding shell associated with the wind outflow it 
is convenient to separate the whole post-shock flow on to four distinct zones, which are schematically shown on the left in
Fig \ref{fig:schem} (see also Figure 5 of \cite{sharma2013}). The zones described below are shown in the schematic diagram as being
separated by dotted lines, from top to bottom: 
{\it i)} the viscous layer where kinetic energy of the inflowing gas is transformed partly in to 
thermal energy, and the entropy 
grows to the post-shock value \citep{zeld1966},  {\it ii)} the radiative 
relaxation layer (RRL) where radiative losses 
lead to the formation of a dense shell, {\it iii)} thin dense shell restricted from the bottom by {\it  iv)} the 
still hot low-density gas formed by %\sout{initial times of shock propagation} 
the termination shock. 
Dissipation due to viscous forces brings the gas into a new high-temperature 
state and vanishes beyond the viscous layer. The thickness of this layer is 
determined by the viscosity, but for strong shock it can be as thin as a few 
free-path length of gas particles \citep{zeld1966}. In numerical models its 
thickness is always unresolved as this thickness is equivalent to a few times the mean-free path. The rate of energy loss 
due to radiation losses in the second zone depends on post-shock temperature 
and density and becomes important when the dynamical time becomes comparable 
to the cooling time.

We show in Appendix C that the density in the cool/dense shell (region {\it iii}) can
be larger than the ambient density by a factor of a few to $\sim 100$. Magnetic field
in the ISM prevents the shell density from becoming exceedingly large.
We characterise the density in the shell as,
\be
n_{sh} = 10 \,  \eta_{10} \, n_{amb} \,,
\ee
where $n_{amb}=n_0e^{-z_+/z_0}$.
Given this density jump, the shell thickness ($\bigtriangleup \rm r_{sh}$) 
can be found by equating the total swept up mass to the shell mass,
\begin{equation}
\int_0 ^{z_+} 4 \pi z^2 n_0 e^{-z/z_0} dz \approx 4\pi z_+^2{\bigtriangleup \rm r_{sh}}n_{sh}\,, 
\end{equation}
which gives the value of $\bigtriangleup \rm r_{sh}$.
Here we have assumed the shell to be spherical,
which is a reasonable approximation since the spherical shape of superbubble shell 
is roughly maintained until it reaches a few times ($2 \textendash 3$) the 
scale height. The shell thickness is shown in the left-most panel of Figure \ref{fig:delr_Av_zp} for different cases. The lower set of
curves are for $\eta_{10}=10$ and the upper set of curves, for $\eta_{10}=1$.

Although the radius and thickness increases with time, since the density in a stratified medium decreases with height,
the total column density in the shell does not increase monotonically. It rises until the shell reaches the scale height
and then decreases. We show the evolution of the total opacity in the shell (proportional to the column density) in the
middle panel of Figure \ref{fig:delr_Av_zp}. The visual extinction is related to the column density as 
$A_v \approx (N_H/1.9 \times 10^{21} \, {\rm cm}^{-2})$ \citep{bohlin1978}.
%\ccr{(Bohlin \etal 1978).}
We find that the maximum visual extinction $A_v$ lies in the range of 
$10\hbox{--}20$, which is attained at roughly the scale height. Note that the value of $A_v$ does not depend on the value of density 
jump.

%%%%%%%%%%%%%%%%%%%%%%%%%%%%%%%%%%%%%%%%%%%%%%%%%%%%%%%%%%%%%%%%%%%%%%%%%%%%%%%%%%%%%%%%%%%%%%%%%%%%%

The temperature in the shell is determined by the balance of heating and cooling processes, which we consider next.

\subsection{Heating and cooling processes in the shell}
\label{heat}
The physical state and ionization structure of the shell resembles the so-called photo-dissociation region (PDR)
considered in the literature \citep{Hollenbach1997}.  Going outward from the central region, beyond the ionized gas, one
first encounters a region of neutral atoms, after which there are regions whose ionization structure is dominated by
the influence of FUV photons on different trace elements, beginning with carbon. Here we focus on the region
where CO/H$_2$ are produced in significant quantity. According to
\cite{wolfire2010} (their eqn 21), the visual extinction $A_v$ corresponding
to CO and H$_2$ are comparable, and is given by,
\be
A_v ({\rm mol})\approx 0.1 \ln \Bigl [ 3.3 \times 10^7 (G_0/1.7 n)^2 +1\Bigl ] \,,
\label{eq:Av_CO}
\ee
where $G_0$ is the FUV photon density (number of photons per unit volume) obtained from Starburst99 code.

The rightmost panel of Figure \ref{fig:delr_Av_zp} shows the value of $A_v ({\rm mol})$ corresponding to the molecular region that we will focus on, 
calculated with eqn \ref{eq:Av_CO}. Recall the rise and decline in the FUV luminosity of the central OB association (see Fig \ref{fig:lum_comp}). This evolution in the FUV luminosity makes the value of $A_v({\rm mol})$ also rise and then decline (through 
the term $G_0$), as seen Fig \ref{fig:delr_Av_zp}. For the cases of small scale height (solid lines), the shell density decreases
rapidly with height, increasing the value of $A_v({\rm mol})$. 

The corresponding ionization fraction of the free carbon ions $x_{C^+}\equiv n_{C^+}/n_C$ in the carbon dominated region is given by \citep{tielens-book}
\bea
{1-x_{C^+} \over x_{C^+}^2} &\approx& 3.3 \times 10^{-6} \Bigl ( {n \over 10^4 \, {\rm cm}^{-3}} \Bigr )\, \Bigl ( { T \over 300 \, K} \Bigr )^{-0.6}
\nonumber\\ && \times
\Bigl ( { G_0 \over 10^4} \Bigr )^{-1} \, \exp [ 2.6 A_v] \,.
\label{ionization}
\eea
The ionization fraction therefore is given by $x_e=A_C  x_{C^+} $, where $A_C=1.4 \times 10^{-4}$ is the carbon abundance for solar metallicity.
The ionization fraction depends on temperature weakly, and is roughly given by $x_e\approx A_C$, since $x_{C^+}\approx 1$, from
eqn \ref{ionization}.

The dominant heating process in this region of the shell is photoelectric (PE) heating, and it cools through radiation. Given the large
density of this region, the temperature is likely to be in the range of $\sim 10\hbox{--}20$ K. We show in Appendix D and E with detailed calculation the values of the ionization fraction and the equilibrium temperature, for different shell densities, and for different cases of $n_0$ and $z_0$.
The results of the calculation confirms that the ionization fraction $x_e \approx 1.4 \times 10^{-4}$ 
and that the equilibrium temperature is of order $\sim 10\hbox{--}20$ K 
(which, to be specific, we approximate as $15$ K). 
We assume thermal equilibrium to calculate the shell temperature and $x_e$, as heating and cooling time scales of the shell are much shorter than the 
dynamical time throughout the evolution of the shell.
In the following calculations for molecule formation we adopt these values of $x_e$ and $T$.

%%%%%%%%%%%%%%%%%%%%%%%%%%%%%%%%%%%%%%%%%%%%%%%%%%%%%%%%%%%%%%
\section{Molecule formation  and dissociation}
\label{mol_form_destruction}
While the observations of molecular outflows have shown that the velocity and the momentum of 
the molecular component
is in rough agreement with the expectations from a star formation driven outflow, the existence of 
molecules in the outflowing
gas is not trivially explained, for the following reasons. The molecular 
component that is seen at a few hundred pc can either be formed in the 
outflowing gas, or can be a residue of the molecules entrained from the parent
molecular clouds, whatever has survived the destruction process while the shell has evolved and has been shocked. 
In this section, we study the molecule formation and destruction processes in detail.
The two important mechanism for the dissociation of molecules 
are the photo-dissociation and the collisional dissociation.

\subsection{Formation and destruction of molecules in the shell} 
\label{mol_net}
Consider the case of a shell  propagating outward, and being 
irradiated by Lyman-Werner band photons from the OB stars responsible for the outflow in the first place. 
%\subsection{Estimates of molecular mass, speed and size}
We consider the formation and dissociation of molecular hydrogen as hydrogen is the most abundant element.
Photons in the Lyman-Werner band ($11.2 \textendash 13.6$ eV) are responsible for the dissociation of hydrogen 
molecule.  The net rate of formation of molecular hydrogen, for a gas density $n$, is given by,
\begin{eqnarray}
 {{dn_{H_2}} \over dt} &=& \mathcal{R}_f\, n \, n_{HI} -\mathcal{R}_{d,thin} f_{\rm dust,H_2} \, f_{\rm shield, H_2}\,n_{H_2} \,\nonumber\\
 &&- k_{D}\,n\,n_{H_2}\, ,
 \label{eq:mol_formation_rate}
\end{eqnarray}
where $n_{H_2}$ is the number density of $H_2$ molecules. As mentioned above, $\mathcal{R}_{d,thin}$ 
depends on the
radiation field, and therefore on the distance of the shell from the centre and the Luminosity in the 
Lyman-Werner band. 
The formation rate $\mathcal{R}_f \approx 3 \times 10^{-18} T^{1/2}$ cm$^3$ s$^{-1}$, 
for solar metallicity
\citep{hollenbach1979}.  The density and temperature
refers to the shell density and equilibrium temperature calculated in 
Appendix \ref{thermal_equil}.

$\mathcal{R}_{d,thin}=3.3 \times 10^{-11} \, G_{LW}$ s$^{-1}$ is 
the photo-dissociation rate in optically thin gas \citep{draine1996}, $G_{LW}$ 
being the strength of the radiation field in units of Habing field. 
The factor $f_{\rm dust,H_2}$ takes into account the effects of dust extinction 
and $f_{\rm shield, H_2}$, that of $H_2$ self-shielding.  
We calculate $G_{LW}$ similar to the equation \ref{eq:UV_field} with the replacement of 
S$_{FUV}$ by S$_{LW}$, albeit without the extinction factor.

For the shielding due to dust, we have,
\be
f_{\rm dust,H_2}=\exp (-3.5 A_v ({\rm mol})) \,.%A_v={N_H \over 1.9 \times 10^{21} \, {\rm cm}^{-2}} \,,
\label{f1}
\ee
%where the second relation follows from Bohlin \etal (1978). Here $N_H=N_{HI}+N_{HII}+2N_{H_2}$. 
For the self-shielding
due to molecules, we use the fit given by Draine \& Bertoldi (1996):
\be
f_{\rm shield, H_2}={0.965\over (1+x/b_5)^2} +{0.035 \over \sqrt{1+x}} \exp \Bigl (-8.5 \times 10^{-4} \sqrt{1+x} \Bigr ) \,,
\label{f2}
\ee
where $x=N_{H_2}/(5 \times 10^{14}$ cm$^{-2})$, $b_5=b/(10^5$ cm s$^{-1})$, $b^2=kT/m_H$, 
being the Doppler broadening parameter.

The collisional dissociation is another important dissociation process for 
destruction of molecules. This process crucially depends on the shell temperature 
and density. The dissociation rate coefficient ($k_D(n,T)$) is given by \citep{lepp1983},
\begin{equation}
\log k_D(n,T)=\log k_H-\log(k_H/k_L)/(1+n/n_{cr})\,,
\label{eq:coll_diss}
\end{equation}
where $k_H (T)$, and $k_L(T)$ are the dissociation rate coefficients for the 
high, and low density limits respectively given in Table 1 of \cite{lepp1983}. 
The critical density $n_{cr}$ depends on temperature %as,
and is calculated using equation 6 of \cite{lepp1983}).

%\subsection{Formation and destruction time scales}
The formation time scale of molecules is given by,
\be
t_{\rm form}\approx (\mathcal{R}_f \, n)^{-1} \approx 10^{-2} \, {\rm Myr} \, \Bigl ( {n \over 10^3 \, {\rm cm}^{-3}}\Bigr ) ^{-1} \, \Bigl ( {T 
\over 100 \, {\rm K} } \Bigr ) ^{-1/2} \,,
\ee
The photo-dissociation time scale of molecules is given by,
\begin{eqnarray}
t_{\rm dest}&\approx &7 \times 10^{-7} \, {\rm Myr} \, \Bigl ( {S_{LW} \over 10^{53} \, s^{-1}} \Bigr ) ^{-1} \,
\Bigl ({r \over 100 \, {\rm pc}} \Bigr ) ^2 \, \nonumber\\
&&\,\,\,\,\, \times (f_{\rm dust,H_2} \, f_{\rm shield, H_2})^{-1} \,.
\label{eq:t_dest}
\end{eqnarray}
In general, the collisional dissociation time scale is much longer than the photo-dissociation 
time scale, given the low temperature
of the dense shell.

These time scales should be compared with the dynamical time scale of the dense shell. 
We found that
the formation time scale becomes comparable to or shorter than the dynamical 
time scales, when the
shell size exceeds $\sim 200$ pc, signalling the onset of molecule formation.
%%%%%%%%%%%%%%%%%%%%%%%%%%%%%%%%%%%%%%%%%%%%%%%%%%%%%%%%%%%%%%%%%%%%%%%%%%%%%%%%%%%%%%%%%%%%%

\begin{figure}
\centerline{
\epsfxsize=0.55\textwidth
\epsfbox{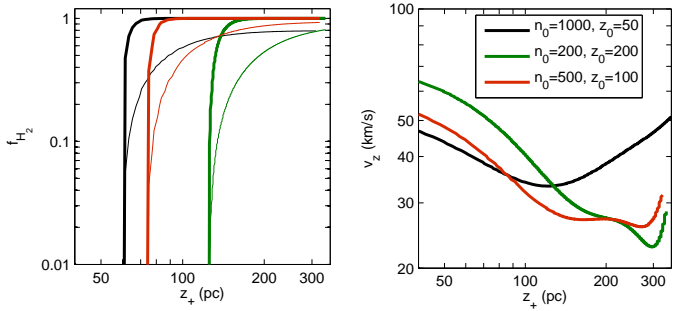}
}
\caption{Evolution of molecular fraction (left), 
and the bubble shell velocity (right) with the size of the superbubble 
shell, for $N_{OB}=10^5$, for  different $n_0$, $z_0$, and $\eta_{10}$ 
cases. The thick, and thin lines correspond to $\eta_{10}=10$, 1 respectively. All the calculations of molecule formation and dissociation are performed in the dense 
superbubble shell after it crosses the D-type ionisation front.
}
\label{fig:mol_r1}
\end{figure}

\begin{figure}
\centerline{
\epsfxsize=0.55\textwidth
\epsfbox{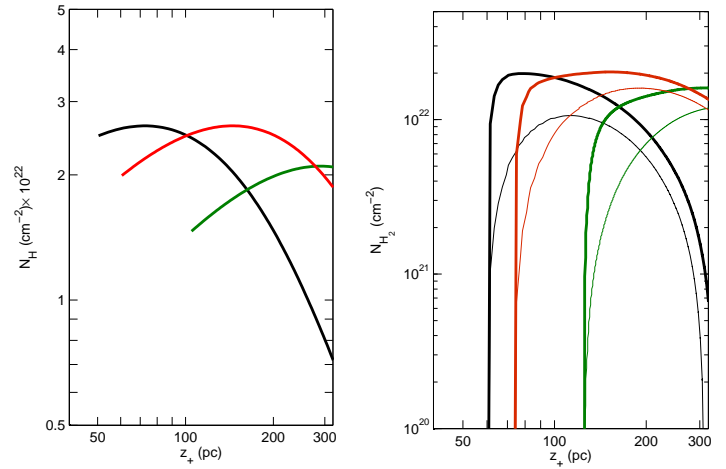}
}
\caption{Evolution of the total hydrogen column density (left) 
and molecular column density (right), with the size of the 
superbubble shell, 
for $N_{OB}=10^5$, for different $n_0$, $z_0$, and $\eta_{10}$ cases. 
The details of the line-styles, and line-colours for different parameters are 
mentioned in the caption of the figure \ref{fig:delr_Av_zp}.}
\label{fig:NH_NH2}
\end{figure}

\subsection{Results}
\label{results}
Figure \ref{fig:mol_r1} (left panel) shows the resulting  
molecular fraction as a function of shell distance for $N_{OB}=10^5$,
for the fiducial cases.  The right panel shows the corresponding velocity 
of the shell.

There is an abrupt jump in the value of the molecular fraction 
($f_{\rm H_2}$) 
from a small value of 
$10^{-6}$ to a maximum value of 1 for large value of $\eta_{10}$ (=10) whereas for the 
case of small
$\eta_{10}$(=1), $f_{\rm H_2}$ increases slowly to 
a maximum value of $0.8\textendash0.9$ for 
all the n$_0\textendash$z$_0$ cases. %For large value of $\eta_{10}$, 
%the temperature is  $\sim 10$ K, favouring molecule formation.
The sharp rise in the molecular fraction corresponds to the epoch of the shell  crossing
 the ionization front, as discussed in \S \ref{thresh}.

The velocity of the superbubble shell for all the n$_0\textendash$z$_0$ cases at $\sim 100\textendash300$ pc
 is $\sim 10-40$ km/s, which matches with the velocities of the molecular outflows as 
 seen in observations, particularly of NGC 253 \citep{bolatto2013}.

Figure \ref{fig:NH_NH2} shows the 
total hydrogen column density and the corresponding molecular column densities, in the left and right
panels respectively. The total column density plots have been explained earlier in the context of total
visual extinction in Figure \ref{fig:delr_Av_zp}. The H$_2$ column density first rises and then falls due
to decreasing shell density and column density. It falls more rapidly in the case of small scale height when
the gas density rapidly decreases with height.

%%%%%%%%%%%%%%%%%%%%%%%%%%%%%%%%%%%%%%%%%%%%%%%%%%%%%%%%%%%%%%%%%%%%%%%%%%%%%%%%%
\begin{figure}
\centerline{
\epsfxsize=0.52\textwidth
\epsfbox{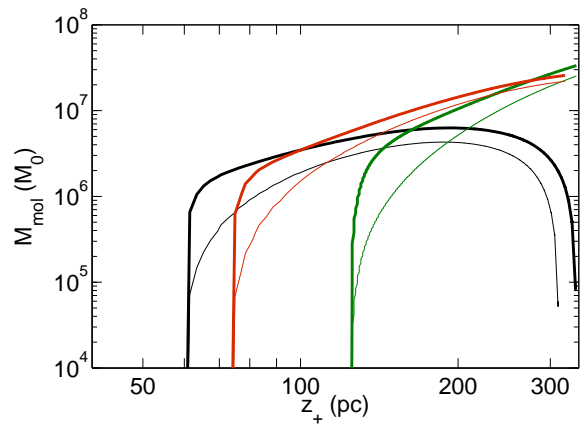}
}
\caption{The evolution of  molecular mass  with the size of the superbubble 
shell, 
for $N_{OB}=10^5$, for three different n$_0\textendash$z$_0$ cases, and 
for the two different values of $\eta_{10}$ (1, 10). Refer to figure 
\ref{fig:delr_Av_zp} for the details of the different line-styles, and line-colours. The molecular 
mass is the integrated mass over the molecular region of the shell.  
}
\label{fig:mol_r2}
\end{figure}
%%%%%%%%%%%%%%%%%%%%%%%%%%%%%%%%%%%%%%%%%%%%%%%%%%%%%%%%%%%%%%%%%%%%%%%%%%%%%%%%%%%%%%%%%%%%%%%%%%%%%%%%%%%%%%%%%%

Figure \ref{fig:mol_r2} shows the evolution of the total molecular mass (integrated over the shell) for different parameters. 
The molecular mass
is found to be in the range of $\sim 10^7\hbox{--}10^8$ M$_\odot$) at $\ge 200$ pc, in the three fiducial
cases, consistent with observations. 
As both the dissociation rates (photo-dissociation, and collisional dissociation) are 
much smaller compared to the formation rate of molecules, for most cases  the molecular mass does not decrease as the 
radius of the superbubble shell increases-- rather its rate of increase may slow down, particularly in the case of disks with small scale
heights. 
For the cases of large mid-plane density and small scale height, the molecular mass decreases, as the column density of the shell
decreases with increasing height.
We note that at larger radii, when the shell crosses a few scale heights, the 
shock-heating can be an important mechanism (as the velocity goes up till $\sim 50\textendash100$ km/s, and also the 
density of the shell reduces) to destroy the molecules in the dense shell, and thus reducing the molecular mass integrated over the shell.
This aspect is beyond the scope of the present paper.

%%%%%%%%%%%%%%%%%%%%%%%%%%%%%%%%%%%%%%%%%%%%%%%%%%%%%%%%%%%%%%%%%%%%%%%%%%%%%%%%%%%%%%%%%%%%%%%%%%%%%%%

\begin{figure}
\centerline{
\epsfxsize=0.52\textwidth
\epsfbox{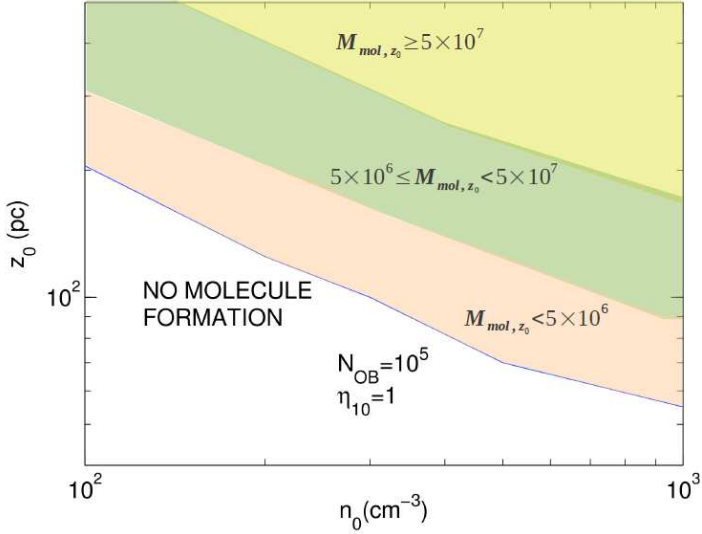}
}
\caption{Regions in the parameter space of $n_0$ and $z_0$ that can give rise to molecular mass of different ranges in shells
triggered by star formation activity, with $N_{\rm OB}=10^5$, and with $\eta_{\rm 10}=1$. Corresponding CO luminosities are also
indicated.}
\label{fig:mass_lum_contour}
\end{figure}

The important parameters that can be used to compare with observations are the molecular mass and the 
length scale of the molecular
shell (as well as its speed). We wish to determine the ranges in the combination of gas density and 
scale height (for the given $N_{OB}=10^5$)
that can give rise to molecular outflows with a certain molecular mass and radius of the shell. 
We consider the molecular mass attained at the time of crossing the disk scale height. We show in Figure \ref{fig:mass_lum_contour} the regions in the parameter space of $n_0$ and $z_0$ that correspond to molecular masses
(calculated at the scale height) of different ranges. 
The lower blank portion corresponds to the lower limit of Figure \ref{fig:threshold}.

%%%%%%%%%%%%%%%%%%%%%%%%%%%%%%%%%%%%%%%%%%%%%%%%%%%%%%%%%%%%%%%%%%%%%%%%%%%%%%%%%%%%%%%%%%%%%%%%%%%%%%%%%%%%%%%%%%%%%%%%%%%%%%%%%%%%%%%
%%%%%%%%%%%%%%%%%%%%%%%%%%%%%%%%%%%%%%%%%%%%%%%%%%%%%%%%%%%%%%%%%%%%%%%%%%%%%%%%%
%%%%%%%%%%%%%%%%%%%%%%%%%%%%%%%%%%%%%%%%%%%%%%%%%%%%%%%%%%%%%%%%%%%%%%%%%%%%%%%%%%%%%%%%%%%%%%%%%%%%%%%%%%%%%%%%%%

\section{Discussion} 
\subsection{Previous studies}
Despite a rather long history of numerical simulations of galactic winds (see the seminal papers \cite{suchkov1994,suchkov1996}), only one paper with theoretical analysis of galactic winds published to date appears to be relevant for the formation of molecular outflows. \cite{thompson2016} described a {\it steady-state} model of a galactic wind with a detailed description of the thermal processes, such as radiative cooling and thermal conduction, along with a qualitative analysis of thermal instability and convection in the expanding gas. Their study focuses, however, on generic features of such a flow, connecting the central energy and mass source and a distant circumgalactic gas. The inner boundary conditions here are set 
 at  $R=0.3$ kpc from a central energy and mass injection source as described by \cite{chevalier1985}, which is actually comparable to the sizes of regions where molecules are already seen in galactic winds. Moreover, gas density ($n\leq 1$ cm$^{-3}$) and temperature ($T\geq 2\times 10^3$ K) in the outflow always remain in the range that is unsuitable for the formation of molecules.

A more plausible scenario for the observed molecular outflows,  and the one that we favour,
is the cooling
of the shocked gas in the radiative outer shock. \cite{thompson2016} (see also 
\cite{sarkar2016}) only consider multiphase cooling in the mass-loaded galactic 
outflow and not the dense shell.
 {\it Steady-state} flows are smooth, showing only a very weak  increase in the gas density at $z\simgt 10$ kpc (see Fig. 2 of \cite{thompson2016}). In our {\it non-steady} outflow, the formation of a shock in the very beginning (at scales of a few pc) is an essential event which makes the gas cool down quickly to $T\simlt 100$ K within  a few Myr or less. At this stage, the ambient gas density is quite high and therefore the post-shock density and temperature are in the range $n>10^3$ cm$^{-3}$ and $T\simlt 10-30$ K where molecules can efficiently form. Moreover, these conditions are suitable for line emission of CO and HCN/HCO$^+$.

\cite{girichidis2016} simulated the effect of SNe driven outflows and studied the effects of clustering and frequency of SNe. Their simulations showed that molecules formed at vertical distances less than the disc scale height (their Figure 2). However, their focus was on the velocity dispersion of different phases of gas in a typical disc galaxy, and not on the dynamics and chemistry of expanding shells triggered by large OB associations, as have been observed in starburst nuclei.

We also note that \cite{zubovas2014} considered the formation of molecules in AGN driven outflows. However, their model is more relevant for the formation of molecules in outflows with speed $\sim 1000$ km s$^{-1}$ and with mass outflow rates of $\sim 1000$ M$_\odot$ yr$^{-1}$,
and is different in scope and nature than the small scale outflows in starburst nuclei with smaller speeds.

\subsection{Comparison with observations}
It is readily seen from Figure \ref{fig:mol_r2} that in a wide range of parameter space ($n_0,z_0$), the model predictions are consistent with the observed molecular outflows. Moreover, numerical models  reveal the range of physical parameters under which starbursts generate powerful high-mass molecular outflows. In particular, it is clear from Figure \ref{fig:mol_r2} that for small gas scale heights $z_0$ the expanding molecular layers widen, thereby decreasing the column density due to mass conservation $N_{\rm H_2}\propto (z_0/z)^{2}$ with nearly constant molecular fraction. The decrease in $N_{\rm H_2}$ is faster  in models with smaller $z_0$. As a result, the maximum molecular mass in such models remains relatively low $M_{\rm mol}\sim 10^6~M_\odot$, while larger scale heights can give rise to increasing molecular mass beyond $\sim 300$ pc. Unfortunately, observations of molecular winds are yet too few, and such an interrelation between density profiles in gas preceding the starburst cannot be inferred from the available observational data.

\begin{figure}
\centerline{
\epsfxsize=0.52\textwidth
\epsfbox{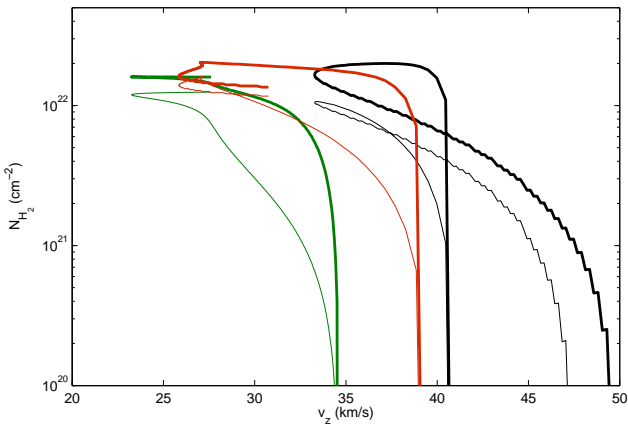}
}
\caption{Molecular column density is plotted against the expansion velocity. The line styles correspond to the same cases as in Figure
\ref{fig:delr_Av_zp}.}
\label{fig:molmass-vel}
\end{figure}

In general, our calculations show that competitive processes governing the thermodynamics and chemical kinetics of the shell, {\it viz}, the heating and ionization by Lyman continuum from the underlying nuclear stellar cluster on one side, and from the shock front on the other, and an additional effect from magnetic pressure -- determine the possibility to form and expel high velocity molecular gas into the outflow. It is seen from Figure \ref{fig:mol_r1} and \ref{fig:NH_NH2} that in general for a given $n_0$-$z_0$ pair, the molecular outflow may show distinct paths on the $N_{\rm H_2}-v$ plane, as shown separately in Figure \ref{fig:molmass-vel}.

We find that the H$_2$ column density grows steeply at a nearly constant velocity when the outflowing shell is small. Then, at higher $z$ the expansion velocity decreases while the H$_2$ column density continues to grow. At a certain point close to the breakout level, the column density starts to decrease and the velocity increases -- this path is clearly seen for the case of small scale height of $z_0=50$ pc (black solid line). For larger values of $z_0$,  the curve $N_{\rm H_2}(v)$ shifts left (decreasing velocity), and the loop occurs at a lower velocity, $v_{\rm min}\propto \rho_0^{-1/3}z_0^{-2/3}$. Even though we show results for a fixed numbers of massive stars $N_{\rm OB}$ in the cluster, this relation allows to predict that in general the scaling would be $v_{\rm min} \propto N_{\rm OB}^{1/3}\rho_0^{-1/3}z_0^{-2/3}$, as follows from a simple wind model in an exponential density profile.

It is worth noting that in realistic conditions the molecular layer is expected to disintegrate due to Rayleigh-Taylor instability with fragments moving outwards nearly ballistically, with the velocity close to the minimum velocity at the loop. In general, the velocity range spanned by  the paths agrees with observations. One can therefore expect that the range of observed velocities in molecular outflows relates to the central gas density, its scale height and likely the number of massive stars, and may help to constrain them.

\subsection{Off-centered shells}
\begin{figure}
\centerline{
\epsfxsize=0.52\textwidth
\epsfbox{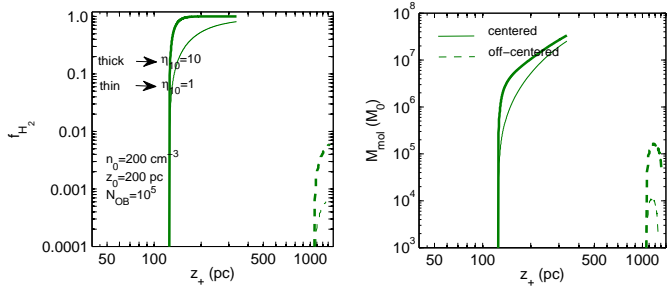}
}
\caption{Molecular fraction and  molecular mass  for
an off-centered superbubble (centered at $z'=z_0$) are compared with the case of superbubbles located at the mid-plane, for $n_0=200$ cm$^{-3}$ and $z_0=200$ pc.}
\label{fig:offcenter}
\end{figure}

We have considered the OB associations that trigger the expanding shells to be located at the mid-plane for simplicity of calculations. However, in reality it could be situated at some height $z\prime \le z_0$, the scale height. Below, we briefly consider the case of off-centered expanding shells. As an extreme case, we show in Figure \ref{fig:offcenter} the case of a superbubble triggered by an association at a height $z\prime=z_0$, and compare the molecular fraction and molecular mass with those of a superbubble located at mid-plane.

In the case of an off-center location of stellar cluster, there are two competing factors: (1) the decreasing column density of the shell and (2) the large distance traversed by the ionization front, because of density stratification. The column density of the shell roughly decreases as (since the accumulated mass is being distributed over the expanding shell),
\be
N_H \approx {n_0 z_0 \over 3} \Bigl ( {z_0\over z_+ -z\prime } \Bigr )^2 \,.
\ee
When the shell crosses the ionization front at $z_I$, the visual extinction should exceed a critical value ($\sim 3$) in order to form a substantial amount of molecules. In other words,
\be
{n_0 z_0\over 3 \times 1.9 \times 10^{21} \, {\rm cm}^{-2}} \Bigl ( {z_0\over z_I -z\prime } \Bigr )^2 \ge 3 \,.
\ee
In this case illustrated above, for $n_0=200$ cm$^{-3}$, $z_0=200$ pc, and $z^\prime=z_0$, the ionization front quickly reaches $z_I \sim 1 $ kpc. Therefore the above condition is not satisfied and molecules do not form in substantial quantity. This estimate could, in principle, put constraints on the height $z^\prime/z_0$ of the stellar cluster that can trigger a molecular outflow.
However, one could argue that such a large OB association (with $N_{\rm OB}=10^5$) is likely to be located close to the mid-plane, where the density is highest, rather than being far above the disk. Another point to note is that the expansion speed of an off-centered expanding shell will  increase monotonically, if $z/z_0 \sim 1$, and will not have any loops as shown in Figure \ref{fig:molmass-vel}.

\subsection{Caveats}
We had set out to understand the dynamical parameters of observed molecular outflows in nearby starburst galaxies, namely, the
length scales and velocities, as well as the corresponding molecular masses. We also sought to understand the possible range
of parameters for producing molecular outflows in highly energetic winds. Our 1-D calculations, with simplified assumptions of 
molecule formation in a spherical shell, show that it is possible to understand the observed outflow sizes ($\ge 50$ pc) and velocities 
($\sim 30\hbox{--}100$ km s$^{-1}$) in the context of superbubbles triggered by starburst activity. The corresponding predicted molecular masses are of order several times $10^6\hbox{--}10^7$ M$_\odot$, also consistent with observations of NGC 253 and NGC 3628. The morphology, dynamics and molecular masses of M82 are admittedly not explained by our simplified model, and therefore we wish to point out various caveats in our calculations.

To begin with, it is not possible to discuss the morphologies of observed molecular outflows with a 1-D calculation, and as a first
step towards understanding this phenomenon our strategy has been to assume a spherical shell. In reality, this shell is likely to fragment,
especially after the shell has broken out of the disk, due to thermal and Rayleigh-Taylor instabilities \citep{roy2013}. Therefore the covering
factor of the shell is likely to be much smaller than unity, allowing radiation and gas to leak through. This is consistent with the observation
of H$\alpha$ radiation from gas far beyond the molecular outflow in M82, for example.
These considerations imply that our estimate of molecular mass is at best approximate, and should be viewed/used with caution. 
As we mentioned in the introduction that the morphology and dynamics of molecular outflows are diverse, and therefore a 
better estimate would have to consider the details of the PDR region in the shell, and possible clumping in it, 
as well as the differences in the formation/destruction processes for different molecules, which are beyond the scope of the present paper.

\section{Summary \& Conclusions}
We summarise our main findings as follows: 
\begin{itemize}
\item We have considered a simple 1-D model of molecule formation in expanding superbubble shells triggered by star formation activity
in the nuclei of starburst galaxies. We have determined a threshold condition (eqn 5) for disk parameters (gas density and scale height) for the formation of molecules in superbubble shells breaking out of disk galaxies. This threshold condition implies a gas surface density of $ \ge 2000$ M$_{\odot}$ pc$^{-2}$, which translates to a SFR of $\ge 3$ M$_{\odot}$ yr$^{-1}$ 
within the nucleus region of radius $\sim  300$ pc, consistent with  observed SFR of galaxies hosting molecular outflows. 
We also show that there is a range in the surface density of SFR
that is most conducive for the formation of molecular outflows, given by $10\le \Sigma_{SFR} \le 50$ M$_\odot$ yr$^{-1}$ kpc$^{-2}$, 
consistent with observations.
\item Consideration of molecule formation in these expanding superbubble shells predicts molecular outflows with velocities $\sim 30\hbox{--}40$ km s$^{-1}$ at distances $\sim 100\hbox{--}200$ pc with a  molecular mass $\sim 10^6\hbox{--}10^7$ M$_{\odot}$, which tally with the recent ALMA observations of NGC 253. 
\item We have considered different combinations of disk parameters and the predicted velocities of molecule bearing shells in the
range of $\sim 30\hbox{--}100$ km s$^{-1}$ with length scales of $\ge 100$ pc are in rough agreement with the observations of molecules in NGC 3628 and M82.
\end{itemize}

\textbf{ACKNOWLEDGEMENT} 

We would like to thank Eve Ostriker for valuable discussions and an anonymous referee for useful comments.
The paper is supported partly (YS) by RFBR (project codes 15-02-08293 and 15-52-45114-IND). YS is also thankful to 
the Grant of the President of the Russian Federation for Support of the Leading Scientific Schools NSh-4235.2014.2 

\footnotesize{

\appendix
\section{Numerical Setup}
\label{num}
In this section, we describe our simulation set up. We use a second-order Eulerian, 
hydrodynamic code (ZEUS-MP code \citep{hayes2006}). We perform a 1D simulation to get superbubble radius, 
and velocity. The equations governing superbubble evolution (see also \cite{roy2013}) are given by,
\begin{equation}
{{d\rho}\over{dt}}=-\rho\nabla{\bf. v}+S_{\rho}(r)\,,
\label{eq:mass_con}
\end{equation}
\begin{equation}
 \rho{{d{\bf v}}\over{dt}}=-\nabla p\,,
 \label{eq:mom_con}
\end{equation}
\begin{equation}
 {{de}\over {dt}}=-q^{-}(n,T)+S_e(r)\,,
\end{equation}
where $\rho$, {\bf v}, and e ($=3p/2$) are the fluid density, velocity, and internal 
energy respectively, p is the thermal pressure of the medium; $S_{\rho}$, $S_e$ are the 
mass and energy source terms respectively, $q^-=n_en_i\Lambda(T)$ is the 
energy loss term due to radiative cooling where $n_e$ ($n_i$), $\Lambda(T)$ are the electron (ion) number 
density, and the cooling function 
respectively. We used \cite{sutherland1993} cooling function for a temperature ranging from
$10^4$ K $\textendash 10^9$ K for solar metallicity; below $10^4$ K we used $\Lambda(T)$ for 
molecular cooling for an electron fraction 
(x$_e=n_e/n_H$) of $\sim 10^{-3}$, guided by the observed the ionization fraction for CNM 
(cold neutral medium), $\sim 10^{-3}\textendash 10^{-4}$\citep{draine2011}. The cooling function 
is also independent of $x_e$, for $10^{-4}<x_e <10^{-2}$, between $10$ K to $10^4$ K. 
We have assumed the initial isothermal ambient medium temperature to be $10$ K.

The initial ambient density is exponentially stratified and is given by, 
$n=n_0\exp(-|r|/z_0)$, where $n_0$, and $z_0$ are the
mid-plane density and the scale height respectively. 
We implement the supenovae (SNe) explosion energy within a small radius 
($r_{in}$) so that a strong shock can move through the interstellar medium 
(ISM) even after the radiative losses \citep{sharma2013}.
The energy source function ($S_e=\mathcal{L}/(4\pi r_{in}^3)$) represents the energy input
 in SNe within radius $r_{in}$, where $\mathcal{L}$ is mechanical luminosity by 
SNe. We implement continuous mechanical luminosity till 
the life-time of 
OB association (10 Myr), $\mathcal{L}= 10^{37} N_{OB}$ erg s$^{-1}$ 
(mechanical luminosities from stellar winds and 
supernova explosions)
as obtained from Starburst99. The mass source function mimics 
the mass injection in SNe. Therefore $S_{\rho}=\dot{M_{in}}/(4\pi r_{in}^3)$, 
where 
$M_{in}$ is the mass ejected by $N_{OB}$. We choose the ejected mass in each supernova 
explosion
to be $\sim 1$ M$_{\odot}$ as superbubble evolution is independent of the mass 
injection rate. We 
assume the injection radius $r_{in}$ to be $2$ pc in all our simulations. 
We also assume the CFL number to be $0.2$ as it is found to be more robust.

We have used spherical ($r$, $\theta$, $\phi$) co-ordinate in all our simulations. We start our 
simulation grid 
from $r_{min}=1$ pc ($<r_{in}$), and the outer boundary extends to $r_{max}$ ($2.5$ kpc). 
To obtain a better resolution in smaller scales, we use 
logarithmically spaced grid points along radial direction. The number of grid points 
between $r_{min}$ and $(r_{min}r_{max})^{1/2}$ are the same as the number of 
grids between $(r_{min}r_{max})^{1/2}$ and $r_{max}$. The $\theta\textendash$boundary runs from 
$0$ to $\pi$, and the $\phi\textendash$boundary extends from $0$ to $2\pi$. We 
adopt the inflow-outflow bounday condition at the inner bounday, and the outflow boundary condition at the 
outer boundary in the radial direction.

In the $r\textendash$direction, we use 512 grid points to calculate the evolution of superbubble shell. 
We show that the time evolution of superbubble shell position is similar for 
different resolutions with a maximum percentage change of 10\% (see appendix \ref{conv}). 
We also show that 
the velocity evolution for different resolutions are also similar (read appendix \ref{conv}). We 
adopt $n_0=200$ cm$^{-3}$, 
and $z_0=200$ pc to be the fiducial case.

%%%%%%%%%%%%%%%%%%%%%%%%%%%%%%%%%%%%%%%%%%%%%%%%%%%%%%%%%%%%%%%%%%%%%%%%%%%%%%%%%%%%%%%%%%%%%%%%%%%%%%%%%
\begin{figure}
\centerline{
\epsfxsize=0.52\textwidth
\epsfbox{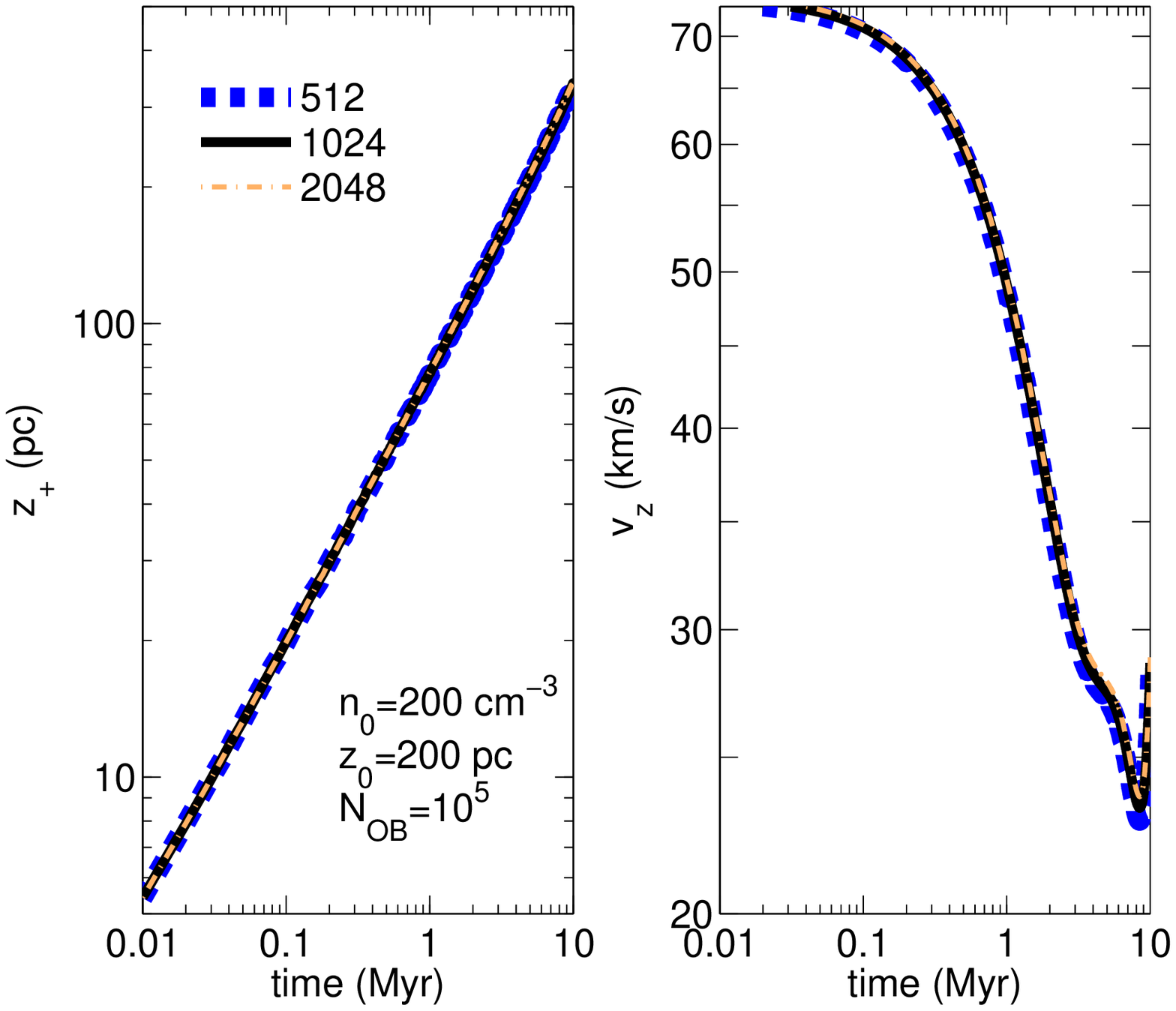}
}
\caption{The time evolution of superbubble shell position and velocity for 
$n_0=200$ cm$^{-3}$, $z_0=200$ pc, and for $N_{OB}=10^5$. The left panel shows the shell position, and the 
right panel represents the shell velocity. The blue-dashed, the black-solid, and the 
light brown dashed-dotted lines are for 512, 1024, 2048 grid points respectively. 
 }
\label{fig:conv_zp_vz}
\end{figure}
%%%%%%%%%%%%%%%%%%%%%%%%%%%%%%%%%%%%%%%%%%%%%%%%%%%%%%%%%%%%%%%%%%%%%%%%%%%%%%%%%%%%%%%%%%%%%%%%%%%%%%%%

\section{Convergence test}
\label{conv}
%\subsection{$z_+$, and $v_z$}
We show the time evolution of the shell position and velocity for three resolutions 
(512, 1024, 2048 grid points), and for $n_0=200$ cm$^{-3}$, $z_0=200$ pc, $N_{OB}=10^5$.
The left panel shows the time evolution of the shell position, and the right panel 
represents the shell velocity.

One can notice that superbubble shell positions for the three resolution 
are comparable. The low resolution 
runs (512, 1024 grid points) show a similar evolution, 
whereas the high resolution case (2048 grid points) varies 
slightly from the low resolution cases with a maximum percentage 
change being of order $\sim 10$\%. One can also 
notice that the velocity evolutions 
show similar 
results as in the case of the evolution of the shell positions (showing that the low resolution 
runs are similar, and a 
maximum percentage change of 10\% for the case of 2048 grid points). 
Therefore, we run all our numerical simulation with 
$512$ grid points with an error of $\sim 10$ \% in both the cases of 
the shell position, and velocity evolution.

\section{Density jump in the superbubble shell}
Since the formation of molecules takes place in the cool/dense shell (region {\it iii}), it is 
important to estimate its density. Let $(\rho_1,~u_1)$ 
and $\rho_3,~u_3$ be the density and velocity of the ISM (region {\it i}) and the shell 
(region {\it iii}), respectively, in the shock rest frame. The conservation of 
mass, momentum, and magnetic flux gives
\bea
\label{eq:mass}
&& \rho_1 u_1 = \rho_3 u_3; \\
\label{eq:momentum}
&& \rho_1 u_1^2 + p_1 + p_{{\rm mag},1}= \rho_3 u_3^2 + p_3 + p_{{\rm mag},3};  \\
\label{eq:mag}
&& B_1 u_1 = B_3 u_3;
\eea
where $p_{1,3}$ ($p_{\rm mag {1,3}}\equiv B_{1,3}^2/[8 \pi]$) is the gas (magnetic) pressure in 
region {\it i,iii}, and $B_{1,3}$ is the field strength in region {\it i,iii}. 
We assume the field lines to be in the shock-plane (this component is 
important in preventing the shell to be compressed to very high densities). All the cooling is 
concentrated in region {\it ii}, and the temperatures 
$T_1$ and $T_3$ correspond to the stable thermal equilibrium temperatures in regions 
{\it i} and {\it iii} (this replaces the energy equation, required to solve for
downstream quantities in region {\it iii}; see Fig. 5 in \citealt{sharma2013}). Lets define the 
compression ratio $r= \rho_3/\rho_1=u_1/u_3=B_3/B_1$. Then, Eqs. \ref{eq:mass}-\ref{eq:mag}, and 
the temperature information gives,
\be
\label{eq:jump}
\frac{r^3}{\beta_1} + \left ( \frac{c_3}{c_1} \right)^2 r^2 - \left( 1+ {\cal M}_1^2 + \frac{1}{\beta_1} \right) r + {\cal M}_1^2 =0,
\ee
where $c_{1,3}^2 \equiv \gamma_{1,3} k_BT_{1,3}/(\mu_{1,3} m_p)$ is the sound speed, 
${\cal M}_1 \equiv u_1/c_1$ is the upstream Mach number, and 
$\beta_1 \equiv 8\pi p_1/B_1^2$ is the upstream plasma $\beta$. Eq. \ref{eq:jump} can be solved 
numerically for various parameters ($\beta_1$, ${\cal M}_1$, 
$c_3/c_1$).

\begin{figure}
\centerline{
\epsfxsize=0.56\textwidth
\epsfbox{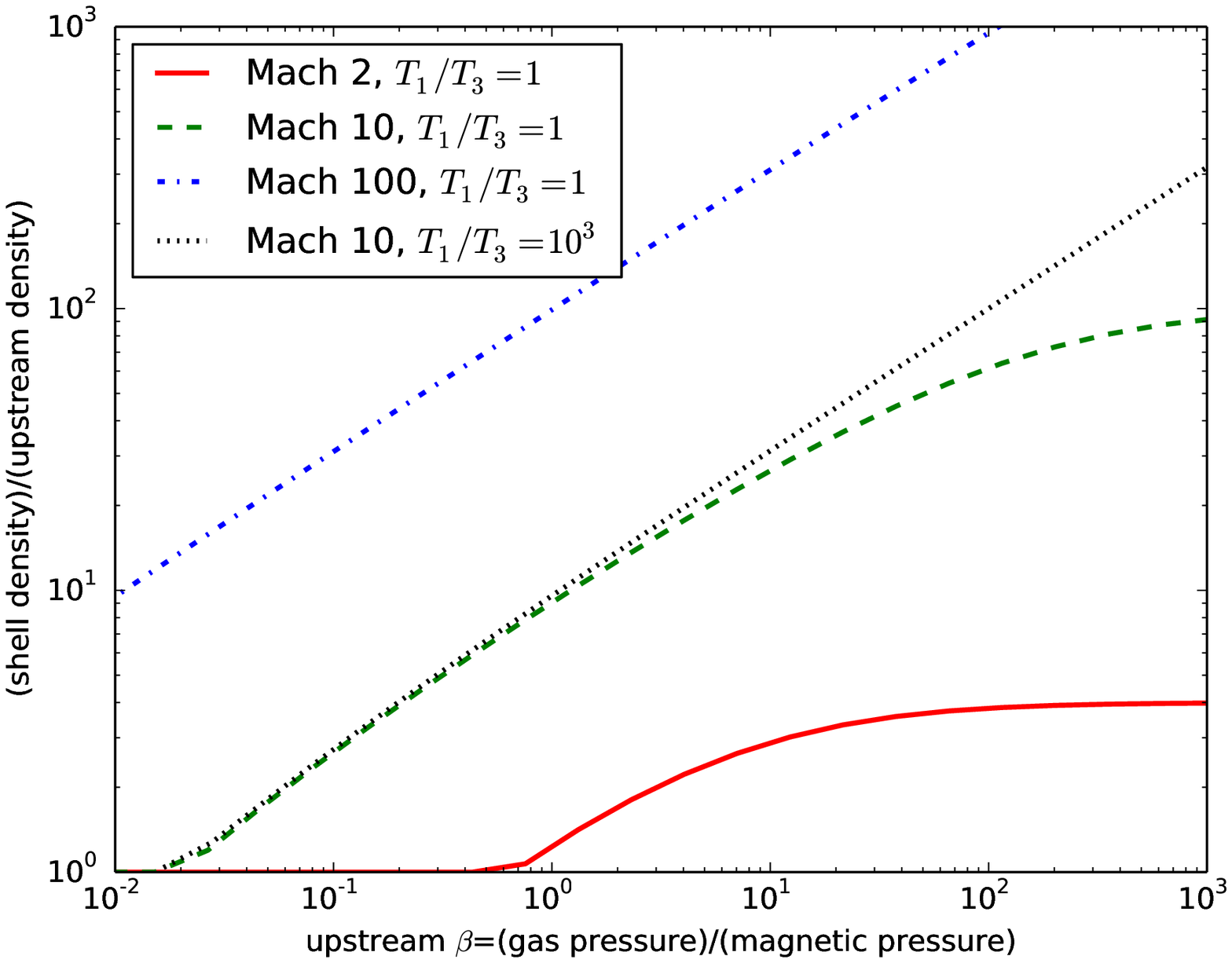}
}
\caption{Compression factor (numerical solution of Eq. \ref{eq:jump}) as a function of upstream 
$\beta_1$ for various values of the upstream Mach number. The influence of the ratio of the 
temperatures in regions {\it 1} and {\it iii} is small in the relevant $\beta_1$ regime.}
\label{fig:compr}
\end{figure}

Figure \ref{fig:compr} shows the compression factor as a function of a reasonable range in upstream plasma 
$\beta$ for three different Mach numbers, 
assuming the same temperature in regions {\it i} and {\it iii} ($c_1=c_3$). As expected, the compression 
factor is larger for a higher Mach number. The compression
factors with a reasonable upstream magnetic field $\beta \sim 1$ is much smaller than the 
hydrodynamic limit ($r \sim {\cal M}_1^2$ as $\beta \rightarrow \infty$). 
A reasonable value for $\rho_3/\rho_1$ for typical ISM $\beta$ is in the range few to 100 
(see also \citealt{smith1993}).

\section{Heating and cooling in the shell}
\label{PE}
The dominant heating process in the molecular region of the shell is photoelectric (PE) heating. We use the PE heating rate 
given by \cite{wolfire2003}, in which they take into 
account the electron-PAH collisions by the term $\phi_{PAH}$ (which takes into account the fraction of PAH). 
At a given density, the heating rate depends on the electron abundance, the diffuse incident 
UV radiation, temperature of the medium, and $\phi_{PAH}$. Therefore, the heating rate per unit volume 
is given by,
\begin{equation}
 n\Gamma_{pe}=1.3\times 10^{-24}n\epsilon G_{FUV} \,{\rm ergs} \, {\rm cm^{-3}}\, {\rm s^{-1}}\,,
\label{eq:heat_pe}
\end{equation}
where n is the density of the hydrogen nucleus, $G_{FUV}$ is the incident UV 
radiation field with the dust-extinction (see equation \ref{eq:UV_field}) in terms of the Habing 
radiation, and $\epsilon$ is the heating efficiency
 given by,
 \begin{eqnarray}
  \epsilon &=& {{4.9\times 10^{-2}} \over {1+4.0\times 10^{-3}(G_{FUV}T^{1/2}/n_e\phi_{PAH})^{0.73}}}\,\nonumber\\
   && + {{3.7\times 10^{-2}(T/10^4)^{0.7}} \over {1+2.0\times 10^{-4}(G_{FUV}T^{1/2}/n_e\phi_{PAH})}}\,,
   \label{eq:heat_eff}
 \end{eqnarray}
\citep{draine2011a}, where $T$ is temperature of the medium, $n_e$ is the electron density. 
The heating rate weakly depends on $\phi_{PAH}$ value varying from 0.25 to 1 
\citep{wolfire2003}. In our calculation, we have assumed the value of 
$\phi_{PAH}\sim 0.5$.

%%%%%%%%%%%%%%%%%%%%%%%%%%%%%%%%%%%%%%%%%%%%%%%%%%%%%%%%%%%%%%%%%%%%%%%%%%%%%%%%%%%%%%%%%%%%%%%%%%%%
\begin{figure}
\centerline{
\epsfxsize=0.52\textwidth
\epsfbox{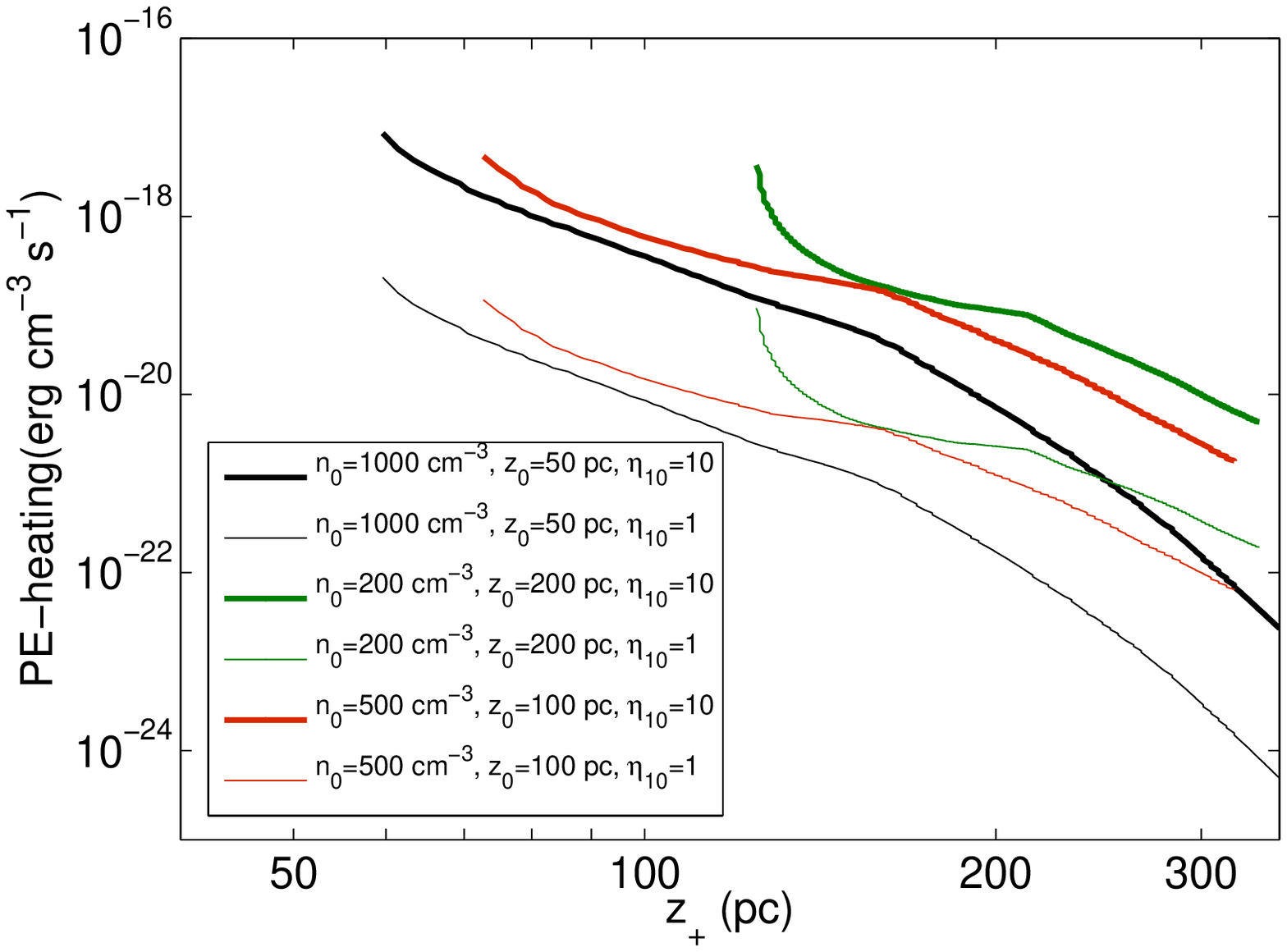}
}
\caption{The time evolution of the heating rate in superbubble shell at the time when its uppermost position 
is $z_+$ , %??} of photo-electric (PE) heating in superbubble shell for 
for three different $n_0$, $z_0$ cases ($n_0=1000$ cm$^{-3}$, $z_0=50$ pc; $n_0=200$ cm$^{-3}$, $z_0=200$ pc, and $n_0=500$ cm$^{-3}$, $z_0=100$ pc), for two $\eta_{10}$ 
cases
and for $N_{OB}=10^5$. 
The thick, and thin solid lines correspond to $\eta_{10}=10$, $1$ respectively. The black, green, and the red lines represent $n_0=1000$ cm$^{-3}$, $z_0=50$ pc; $n_0=200$ cm$^{-3}$, $z_0=200$ pc, and $n_0=500$ cm$^{-3}$, $z_0=100$ pc cases respectively. }
\label{fig:pe_heat_sh_heat_cos_heat_time}
\end{figure}

We assume a central OB association at the centre of 
the galactic 
disk. We use the Starburst99 code to calculate the FUV
($5.4\textendash 13.6$eV) 
photon luminosity ($S_{FUV}$) (see figure \ref{fig:lum_comp} in section 
\ref{n0_z0_limit}) to obtain $G_{FUV}$ as a function of $z_+$ as,
\begin{eqnarray}
G_{FUV}(z_+)&=&{{\bigl(S_{FUV}\exp(-\tau_{FUV})/4\pi z_+^2\bigr)}\over{\bigl(4\times 10^{-14}c/h\nu_{1000 A^{\circ}}\bigr)}}\,\nonumber\\
&=& {{(S_{FUV}\exp(-\tau_{FUV})/4\pi z_+^2)}\over{6 \times 10^7}} \,,
\label{eq:UV_field}
\end{eqnarray}
where $c$ is the speed of light, $\tau_{FUV}$($=\sigma_d A_v({\rm mol})\times 1.87\times 10^{21}$) is the optical depth of the shell for the FUV photons 
for the dust extinction cross-section of $\sigma_d$. We have considered $\sigma_d$ at 1000\angstrom  to be
$\sim 6\times 10^{-22}$ cm$^2$ for dense clouds with reddening parameter 
of R$_v= 5.5$ \citep{draine1996}.
We have used the fact that Habing 
field has an energy density of $4\times 10^{-14}$ erg cm$^{-3}$ 
at $1000$ A$^{\circ}$ \citep{draine1996}.

%The two heating rates are compared 
We show the PE heating in Figure \ref{fig:pe_heat_sh_heat_cos_heat_time} for the two fiducial cases.
We calculate the PE-heating rate once the shell crosses the Str\"omgren sphere 
radius for all the three combinations of n$_0$, z$_0$. We notice 
that PE-heating depends on electron density, and FUV luminosity, 
and the equilibrium shell temperature. On the other hand, the electron density 
depends on the shell density and temperature (as recombination is temperature dependant) which in turn is determined by the heating (PE-heating)
and cooling balance. Thus one needs to solve the equations of ionization and thermal equilibrium 
simultaneously to obtain n$_e$, and 
T$_{shell}$, and to understand their effect on PE-heating rate.
The electron density (n$_e$) has a strong dependence on the shell 
density, thus n$_e$ decreases as the shell density decreases. 
Therefore, the PE-heating rate also drops initially. In all these three n$_0\textendash$ z$_0$ 
combinations the shell radius reaches at $\sim 200$ pc in 
$2\textendash 3$ Myr, when the FUV photon luminosity starts 
dropping drastically, and thus 
we notice kinks 
in the curves of PE-heating rates at $\sim 200$ pc, and PE-heating rate drops after 200 pc due to the 
drop in $S_{FUV}$.

We use the same cooling function as in our simulation for the dynamics of the superbubbles, 
the details of which
are described in \S \ref{num}.

\section{Density and temperature in the dense shell}
\label{thermal_equil}
The heating and cooling time scales in the shell are shorter than the dynamical time scale ($z_+/t$) 
at all times. 
Thus one can assume thermal 
 equilibrium to calculate the shell temperature. 

%%%%%%%%%%%%%%%%%%%%%%%%%%%%%%%%%%%%%%%%%%%%%%%%%%%%%%%%%%%%%%%%%%%%%%%%%%%%%%%%%%%%%%%%%%%%%%%%%%%%%%%
\begin{figure}
\centerline{
\epsfxsize=0.52\textwidth
\epsfbox{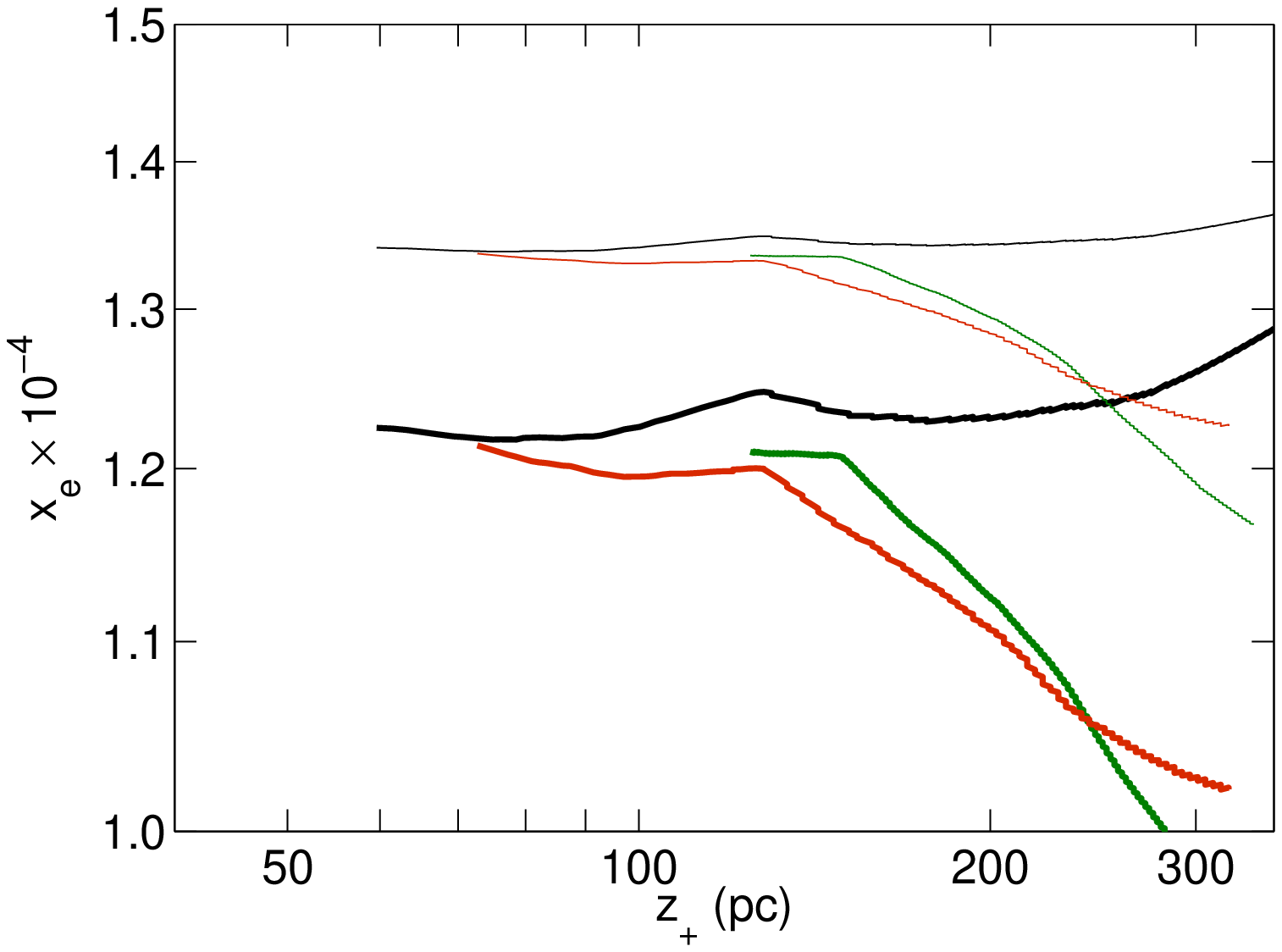}
}
\caption{The evolution of the ionisation fraction 
as a function of the vertical height of the superbubble. 
All the line styles, and line-colours representing different n$_0$, z$_0$, and $\eta_{10}$ 
are mentioned in the caption of figure \ref{fig:pe_heat_sh_heat_cos_heat_time}.
}
\label{fig:den_zp}
\end{figure}
%%%%%%%%%%%%%%%%%%%%%%%%%%%%%%%%%%%%%%%%%%%%%%%%%%%%%%%%%%%%%%%%%%%%%%%%%%%%%%%%%%%%%%%%%%%%%%%%%%

First we show the ionization fraction, total gas density and electron density in the shell in Figure \ref{fig:den_zp}.
The nature of the curve for $n_e$ mimics that of the curve for the total shell 
density, $n_{sh}$, albeit with small differences which show clearly in the plot for $x_e$, the ionization fraction. Here again there are
kinks in the curves at $\sim 200$ pc, and they arise because of the change in the FUV luminosity as mentioned earlier.

%%%%%%%%%%%%%%%%%%%%%%%%%%%%%%%%%%%%%%%%%%%%%%%%%%%%%%%%%%%%%%%%%%%%%%%%%%%%%%%%%%%%%%%%%%%%%%%%%%%%%%
\begin{figure}
\centerline{
\epsfxsize=0.52\textwidth
\epsfbox{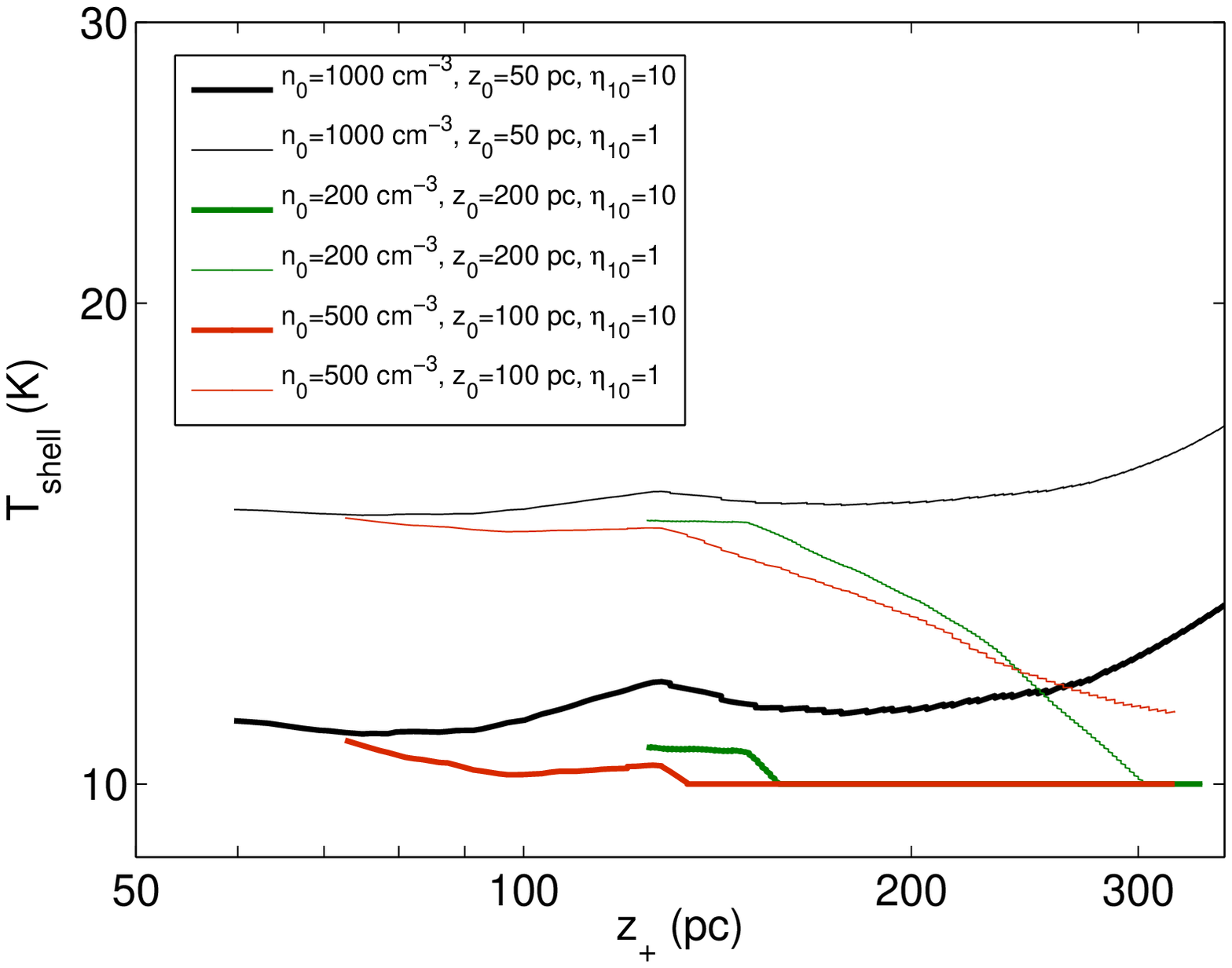}
}
\caption{The equilibrium temperature of the shell is plotted as a function of the shell-radius for 
three difefrent $n_0$, $z_0$ cases for $N_{OB}=10^5$, and 
for $\eta_{10}=1$, 10. 
 }
\label{fig:temp_zp}
\end{figure}
%%%%%%%%%%%%%%%%%%%%%%%%%%%%%%%%%%%%%%%%%%%%%%%%%%%%%%%%%%%%%%%%%%%%%%%%%%%%%%%%%%%%%%%%%%%%%%%%%%%%%

Next we show the equilibrium shell temperature as a function of the position of the 
shell in Figure \ref{fig:temp_zp}, for three different 
combinations of $n_0, z_0$ and for $N_{OB}=10^5$.
In the case of  larger scale height ($200$ pc), the shell temperature initially is $\sim 10\textendash20$ K, and it falls to 
$\sim 10$ K at larger radii. This is owing to the high density in the shell, and consequently, 
greater cooling. 
In the case of smaller scale height, the low density at large heights, the ionization fraction increases and so does PE heating, and thus the shell temperature increases with radius. Again,  kinks arise due to the nature of FUV luminosity evolution.


\begin{thebibliography}{}

\bibitem[\protect\citeauthoryear{Alton, Davies \& Bianchi}{1999}]{alton1999}
Alton, P. B., Davies, J. I., \& Bianchi, S., 1999, A\&A, 343, 51

\bibitem[\protect\citeauthoryear{Bohlin et al.}{1978}]{bohlin1978}
Bohlin, R. C., Savage, B. D., Drake, J. F., 1978, ApJ, 224, 132

\bibitem[\protect\citeauthoryear{Bolatto et al.}{2013}]{bolatto2013}
Bolatto, A. D. \etal 2013, Nature, 499, 450


\bibitem[\protect\citeauthoryear{Bolatto, Wolfire \& Leroy}{2013}]{bolatto2013_Ann}
Bolatto, A. D., Wolfire, M., Leroy, A. K., 2013, Annu. Rev. Astron Astrophys., 51, 207


\bibitem[\protect\citeauthoryear{Chevalier \& Clegg}{1985}]{chevalier1985}
Chevalier, R. A., Clegg, A. W. 1985, Nature, 317, 44

\bibitem[\protect\citeauthoryear{Dale et al.}{2014}]{dale2014}
Dale, J. E., Ngoumou, J.,  Ercolano, B., Bonnel, I. A., 2014, MNRAS, 442, 694

\bibitem[\protect\citeauthoryear{Draine \& Bertoldi}{1996}]{draine1996}
Draine, B. T., Bertoldi, F., 1996, ApJ, 468, 269

\bibitem[\protect\citeauthoryear{Draine}{2011a}]{draine2011a}
Draine, B. T., 2011a, ApJ, 732, 100

\bibitem[\protect\citeauthoryear{Draine}{2011b}]{draine2011}
Draine, B. T., 2011b, Physics of the Interstellar and Intergalactic Medium, Princeton University Press 

\bibitem[\protect\citeauthoryear{Fenech et al.}{2010}]{fenech2010}
Fenech, D., et al., 2010, MNRAS, 408, 607

\bibitem[\protect\citeauthoryear{Freyer, Hensler \& Yorke}{2003}]{freyer2003}
Freyer, T., Hensler, G. \& Yorker, H. W., 2003, ApJ., 594, 888

\bibitem[\protect\citeauthoryear{Garc\'ia et al.}{2013}]{garcia2013}
Garc\'ia, J., Elhoussieny, E. E., Bautista, M. A., Kallman, T. R., 2013, ApJ, 775, 8 

\bibitem[\protect\citeauthoryear{Girichidis et al.}{2016}]{girichidis2016}
Girichidis, P. \etal 2016, MNRAS, 456, 3432

\bibitem[\protect\citeauthoryear{Gould \& Salpeter}{1963}]{salpeter1963}
Gould, R. J., Salpeter, E. E., 1963, ApJ, 138, 393 

\bibitem[\protect\citeauthoryear{Gupta et al.}{2016}]{gupta2016}
Gupta, S., Nath, B. B., Sharma, P., Shchekinov, Y. 2016, arxiv: 1606.09127


\bibitem[\protect\citeauthoryear{Hayes et. al.}{2006}]{hayes2006}
Hayes, J. C., Norman, M. L., Eiedler, R. A., Bordner, J.O., Li, P.S., Clark, S.E., ud-Doula, A., Mac Low, M-M, 2006, ApJS, 165, 188

\bibitem[\protect\citeauthoryear{Heckman et. al.}{2015}]{heckman2015}
Heckman, T. M., Alexandroff, R. M., Borthakur, S., Overzier, R., Leitherer, C. 2015, ApJ, 809, 147


\bibitem[\protect\citeauthoryear{Heckman \etal}{2000}]{heckman00}
Heckman, T. M., Lehnert, M. D., Strickland, D. K., Armus, L. 2000, ApJ, 129, 493


\bibitem[\protect\citeauthoryear{Heckman et. al.}{1990}]{heckman1990}
Heckman, T. M., Armus, L., George, K. M. 1990, ApJS, 74, 833


\bibitem[\protect\citeauthoryear{Hollenbach \& McKee}{1979}]{hollenbach1979}
Hollenbach, D., McKee, C. F. 1979, ApJS, 41, 555

\bibitem[\protect\citeauthoryear{Hollenbach \& Tielens}{1997}]{Hollenbach1997}
Hollenbach, D., Tielens, A. G. G. M. 1997, ARAA, 35, 197

\bibitem[\protect\citeauthoryear{Sarkar \etal}{2015}]{sarkar2015}
Sarkar, K. C., Nath, B. B., Sharma, P., Shchekinov, Y. 2015, MNRAS, 448, 328

\bibitem[\protect\citeauthoryear{Kennicutt}{1998}]{kennicutt1998}
Kennicutt, R, C. Jr. 1998, ApJ, 498, 541

\bibitem[\protect\citeauthoryear{Kompaneets}{1960}]{kompaneets1960} 
 Kompaneets, A. S. 1960, Soviet Phys Dokl., 5. 46 
 
\bibitem[\protect\citeauthoryear{Korolev et al.}{2015}]{korovel2015}  
Korolev, V., Vasiliev E. O., Kovalenko, I. G., Shchekinov, Yu. A., 2015, ARep, 59, 690 



\bibitem[\protect\citeauthoryear{Lepp \& Shull}{1983}]{lepp1983}
Lepp, S., Shull, J. M., 1983, ApJ, 270, 578

\bibitem[\protect\citeauthoryear{Mac-Low et al}{1998}]{mmml}
Mac-Low M.-M., Norman, M. 1998, ApJ, 

\bibitem[\protect\citeauthoryear{Martin}{2005}]{martin2005}
Martin, C. 2005, ApJ, 621, 227

\bibitem[\protect\citeauthoryear{Martin, Keogh \& Mandy}{1998}]{martin1998}
Martin P. G., Keogh W. J. \& Mandy M. E., 1998, ApJ, 499, 793

\bibitem[\protect\citeauthoryear{Nath \& Shchekinov }%
{2013}]{nath2013} Nath, B. B., Shchekinov, Y.  2013, ApJ, 777, L12

\bibitem[\protect\citeauthoryear{Rosen et al. }%
{2015}]{rosen14} Rosen, A., et al.,  2014, MNRAS, 442, 2701


\bibitem[\protect\citeauthoryear{Roy et. al.}{2013}]{roy2013}
Roy, A., Nath, B. B., Sharma, P., Shchekinov, Y., 2013, MNRAS, 434, 3572

\bibitem[\protect\citeauthoryear{Salak \etal}{2014}]{salak2013}
Salak, D., Nakai, N., Miyamoto, Y., Yamauchi, A., Tsuru, T. G., 2013, PASJ, 65, 66

\bibitem[\protect\citeauthoryear{Salas \etal}{2014}]{salas2014}
Salas, P., Galaz, G., Salter, D. Herrera-Camus, R., Bollatto, A. D., Kepley, A. 2014, ApJ, 797, 134

\bibitem[\protect\citeauthoryear{Salak \etal}{2016}]{salak2016}
Salak, D., etal 2016, ApJ, 823, 68

\bibitem[\protect\citeauthoryear{Sarkar \etal}{2016}]{sarkar2016}
Sarkar, K. C., Nath, B. B., Sharma, P., Shchekinov, Y.  2016, ApJL, 818, L1


\bibitem[\protect\citeauthoryear{Sharma et. al.}{2014}]{sharma2013}
Sharma, P., Roy, A., Nath, B. B., Shchekinov, Y., 2014, MNRAS, 443, 3463

\bibitem[\protect\citeauthoryear{Smith}{1993}]{smith1993}
Smith, M. D., 1993, A\&A, 272, 571 

\bibitem[\protect\citeauthoryear{Smith \& Cox}{1993}]{smith1993}
Smith \& Cox 1993, ApJ  

\bibitem[\protect\citeauthoryear{Smith}{2014}]{smith14}
Smith, N., 2014, ARA\&\,A, 52, 487

\bibitem[\protect\citeauthoryear{Strickland et al.}{2004}]{strickland2004}
Strickland, D. K, Heckman, T. M., Colbert, E. J. M., Hoopes, C. G., Weaver, K. A., 2004, ApJ, 606, 829

\bibitem[\protect\citeauthoryear{Suchkov et al.}{1994}]{suchkov1994}
Suchkov, A. A., Balsara, D. S., Heckman, T. M., Leitherer, C. 1994, ApJ, 430, 511


\bibitem[\protect\citeauthoryear{Suchkov et al.}{1996}]{suchkov1996}
Suchkov, A. A., Berman, V. G., Heckman, T. M., Balsara, D. S. 1996, ApJ, 463, 518

\bibitem[\protect\citeauthoryear{Sunyaev \& Strelnitsky}{1993}]{sunyaev}
Sunyaev, R. A., Strelnitsky, V. S., Astron. Rept. 

\bibitem[\protect\citeauthoryear{Sutherland \& Dopita}{1993}]{sutherland1993}
Sutherland, R. S., Dopita, M. A. 1993, ApJS, 88, 253

\bibitem[\protect\citeauthoryear{Thompson et al.}{2016}]{thompson2016}
Thompson, T. A., Quataert, E., Zhang, D., Weinberg, D. H. 2016, MNRAS, 455, 1830

\bibitem[\protect\citeauthoryear{Tielens}{2005}]{tielens-book}
Tielens, A. G. G. M. 2005, {\it The physics and chemistry of interstellar medium}, Cambridge University Press.

\bibitem[\protect\citeauthoryear{Tsai et al.}{2012}]{tsai2012}
Tsai, A.-L., Matsushita, S., Kong, K. H., Matsumoto, H., Kohno, K. 2012, ApJ, 752, 38

\bibitem[\protect\citeauthoryear{Vasiliev et al.}{2015}]{vasiliev2015}
Vasiliev, E. O., Nath, B. B., Shchekinov, Y. 2015, MNRAS, 446, 1703


\bibitem[\protect\citeauthoryear{Walter et al.}{2002}]{walter2002}
Walter, F., Weiss, A., Scoville, N. 2002, ApJL, 580, 21

\bibitem[\protect\citeauthoryear{Weaver \etal} {1977}]{weaver1977}
Weaver R., McCray R., Castor J., Shapiro P., Moore R., 1977, ApJ, 218, 377

\bibitem[\protect\citeauthoryear{Wolfire et. al.}{2003}]{wolfire2003}
Wolfire, M. G., McKee, C. F., Hollenbach, D., 2003, ApJ, 587, 278

\bibitem[\protect\citeauthoryear{Wolfire et. al.}{2010}]{wolfire2010}
Wolfire, M. G., Hollenbach, D. McKee, C. F. 2010, ApJ, 716, 1191

\bibitem[\protect\citeauthoryear{Zeldovich \& Raizer}{1966}]{zeld1966}
Zeldovich, Ya. B., Raizer, Yu. P., Physics of Shock Waves and High-Temperature Hydrodynamic Phenomena, Academic Press 

\bibitem[\protect\citeauthoryear{Zhao et al.}{1997}]{zhao1997}
Zhao, J.-H., Anantharamaiah, K. R., Goss, W. M., Viallefond, F. 1997, ApJ, 482, 186

\bibitem[\protect\citeauthoryear{Zubovas \& King}{2014}]{zubovas2014}
Zubovas, K., King, A. R. 2014, MNRAS, 439, 400

\end{thebibliography}
\end{document}